\documentclass[acmsmall,screen]{acmart}

\AtBeginDocument{%
  }

\setcopyright{acmlicensed}
\copyrightyear{2025}
\acmYear{2025}
\acmDOI{XXXXXXX.XXXXXXX}

\acmConference[Conference acronym 'XX]{Make sure to enter the correct
  conference title from your rights confirmation email}{June 03--05,
  2018}{Woodstock, NY}

\acmISBN{978-1-4503-XXXX-X/2018/06}

\usepackage{graphicx}
\usepackage{xcolor}
\usepackage{subcaption}
\usepackage{multirow}
\usepackage{makecell}
\usepackage{pifont}
\usepackage[most]{tcolorbox} 
\tcbuselibrary{listingsutf8} 
\usepackage{listings}  
\usepackage{caption} 
\usepackage{markdown}
\usepackage{mdframed}
\usepackage{tikz}
\usepackage[utf8]{inputenc}
\usepackage{pifont}
\newcommand{\mymodel}{\textsc{OpenCodeEdit}}
\newcommand{\mydataset}{\textsc{OCEData}}
\newcommand{\mydatasetft}{\textsc{OCEDataFT}}
\newcommand{\mydatasetrand}{\textsc{OCEData-Rand}}
\newcommand{\mydescft}{\textsc{OCE-Desc-FT}}
\newcommand{\mylazyft}{\textsc{OCE-Lazy-FT}}
\newcommand{\mydatasetQwen}{\textsc{OCEData-Qwen3}}
\newcommand{\mydatasetQwenft}{\textsc{OCEData-Qwen3-FT}}
\newcommand{\mydatasetQwenrand}{\textsc{OCEData-Qwen3-Rand}}
\newcommand{\mydatasetDS}{\textsc{OCEData-DS}}
\newcommand{\mydatasetDSft}{\textsc{OCEData-DS-FT}}
\newcommand{\mydatasetDSrand}{\textsc{OCEData-DS-Rand}}
\newcommand{\mydatasetQDft}{\textsc{OCEData-QD-FT}}
\newcommand{\myfiltering}{\textsc{DT-Filtering}}
\newcommand{\canitedit}{\texttt{CanItEdit}}
\newcommand{\commitpack}{\texttt{CommitPack}}
\newcommand{\commitpackft}{\texttt{CommitPackFT}}
\newcommand{\diff}{\texttt{diff}}

\newcommand{\parabf}[1]{\noindent\textbf{#1}}

\definecolor{ggray}{HTML}{eff0f0}
\definecolor{gggray}{HTML}{E8E8E8}
\definecolor{ggggray}{HTML}{BEBEBE}

\newcounter{finding}
\newcommand{\finding}[1]{\refstepcounter{finding}
 	\vspace{1mm}
	\begin{mdframed}[linecolor=gray,roundcorner=12pt,backgroundcolor=gray!15,linewidth=3pt,innerleftmargin=2pt, leftmargin=0cm,rightmargin=0cm,topline=false,bottomline=false,rightline = false]
		\textbf{Finding \arabic{finding}:} #1
	\end{mdframed}
	\vspace{1mm}
}

\newcommand{\distance}{5pt}
\newcommand{\llmboxwidth}{0.90\linewidth} 
\newcommand{\llmboxfontsize}{\footnotesize}      

\definecolor{userHeader}{RGB}{10,57,102}    
\definecolor{userBody}{RGB}{225,238,249}    

\definecolor{assistantHeader}{RGB}{0,120,120} 
\definecolor{assistantBody}{RGB}{224,242,240} 

\definecolor{systemHeader}{RGB}{204,119,34}   
\definecolor{systemBody}{RGB}{255,244,230}    

\tcbset{
  commonbox/.style={
    enhanced,
    breakable,
    boxrule=0.8pt,
    arc=1.5mm,
    left=4mm, right=4mm, top=2mm, bottom=2mm,
    pad at break=0.4mm,        
    frame hidden,             
    width=\llmboxwidth,
    fonttitle=\bfseries,
    coltitle=white,
    before skip=6pt, after skip=6pt,
    fontupper=\llmboxfontsize,
  },
  userbox/.style={
    commonbox,
    title={User},
    colframe=userHeader,
    colback=userBody,
    colbacktitle=userHeader
  },
  assistantbox/.style={
    commonbox,
    title={Assistant},
    colframe=assistantHeader,
    colback=assistantBody,
    colbacktitle=assistantHeader
  },
  systembox/.style={
    commonbox,
    title={System},
    colframe=systemHeader,
    colback=systemBody,
    colbacktitle=systemHeader
  },
  instrExample/.style={
        enhanced,
        colback=white,
        colframe=black,
        arc=3mm,
        height=6.5cm,
        valign=center,
        fonttitle=\bfseries,
        fontupper=\footnotesize
    }
}

\begin{document}

\title{Generating High-Quality Datasets for Code Editing via Open-Source Language Models}

\author{Zekai Zhang}
\email{zhangzk27@mail2.sysu.edu.cn}
\orcid{0009-0001-1028-2924}
\affiliation{%
  \institution{School of Software Engineering, Sun Yat-sen University}
  \city{Zhuhai}
  \state{Guangdong Province}
  \country{P.R. China}
}

\author{Mingwei Liu}
\email{liumw26@mail.sysu.edu.cn}
\orcid{0000-0002-3462-997X}
\affiliation{%
  \institution{School of Software Engineering, Sun Yat-sen University}
  \city{Zhuhai}
  \country{P.R. China}
}

\author{Zhenxi Chen}
\affiliation{%
  \institution{School of Software Engineering, Sun Yat-sen University}
  \city{Zhuhai}
  \country{P.R. China}
}

\author{Linxi Liang}
\affiliation{%
 \institution{School of Software Engineering, Sun Yat-sen University}
  \city{Zhuhai}
  \country{P.R. China}
}

\author{Yuxuan Chen}
\affiliation{%
  \institution{School of Software Engineering, Sun Yat-sen University}
  \city{Zhuhai}
  \country{P.R. China}
}

\author{Guangsheng Ou}
\affiliation{%
  \institution{School of Software Engineering, Sun Yat-sen University}
  \city{Zhuhai}
  \country{P.R. China}
}

\author{Yanlin Wang}
\affiliation{%
  \institution{School of Software Engineering, Sun Yat-sen University}
  \city{Zhuhai}
  \country{P.R. China}
}
\email{wangylin36@mail.sysu.edu.cn}

\author{Dan Li}
\affiliation{%
  \institution{School of Software Engineering, Sun Yat-sen University}
  \city{Zhuhai}
  \country{P.R. China}
}
\email{lidan263@mail.sysu.edu.cn}
\orcid{0000-0002-3787-1673}

\author{Xin Peng}
\affiliation{%
  \institution{School of Computer Science, Fudan University}
  \city{Shanghai}
  \country{P.R. China}
}
\email{pengxin@fudan.edu.cn}

\author{Zibin Zheng}
\affiliation{%
  \institution{School of Software Engineering, Sun Yat-sen University}
  \city{Zhuhai}
  \country{P.R. China}
}
\email{zhzibin@mail.sysu.edu.cn}
\orcid{0000-0001-7872-7718}

\renewcommand{\shortauthors}{Zhang et al.}

\begin{abstract}
  Code editing plays a vital role in software engineering, requiring developers to adjust existing code according to natural language instructions while keeping functionality intact and avoiding unnecessary modifications. However, commit-based datasets commonly used for this task are often noisy, lack diversity, and fail to reflect the style of real-world edit instructions. To address this, we introduce \mymodel{}, an open-source pipeline that leverages multiple LLMs to synthesize realistic code-edit triplets. The pipeline produces both concise ``lazy'' instructions and more detailed ``descriptive'' ones, and applies filtering based on diffs and topics to guarantee data quality and variety. Using this process, we construct \mydatasetft{}, a curated dataset of 20K samples. Fine-tuning three advanced base models on \mydatasetft{} leads to significant performance boosts on the \canitedit{} benchmark, with relative pass@1 improvements ranging from 4.50\% to 20.79\%. Notably, the resulting models achieve performance close to closed-source systems, narrowing the gap to GPT-4 to just 3.54\%, without relying on proprietary resources or manual annotation. 
All datasets, code, and models are publicly released at~\url{https://github.com/zkzhang88/OpenCodeEdit-public-1}
\end{abstract}



\maketitle

\section{Introduction}
\label{sec:intro}
Code editing is the task of transforming existing source code to satisfy a concrete edit specification, typically expressed as a natural-language instruction, producing a modified program that implements the requested change while minimizing unintended impacts on other parts of the system~\cite{lozhkov2024starcoder,shypula2023learning,cassano2024can,zhang2023self}. This task includes bug fixes~\cite{xie2025swe,xia2024automated,moon2023coffee,chen2023teaching}, refactorings~\cite{shirafuji2023refactoring,cummins2024don}, API migrations~\cite{ziftci2025migrating,almeida2024automatic,zhou2023hybrid}, performance optimizations~\cite{shypula2023learning,wei2024improving,ouyang2025kernelbench}, and feature additions~\cite{guo2025omnigirl,gupta2023grace,li2025fea,deng2025nocode}; unlike pure code generation, editing requires understanding program context, cross-file dependencies, and intended semantics to make precise, often localized changes. Code editing is common in software maintenance and dominates much of enterprise development work~\cite{liu2024coedpilot}. Empirical studies on open-source projects have shown that ``edits'' account for approximately 70\% of all commits~\cite{nguyen2013study}. Performing these edits manually is time-consuming and error-prone, making automation crucial for improving efficiency and developer productivity~\cite{labash2024res,liu2024coedpilot,gupta2023grace}.

Recent advances in Large Language Models (LLMs) have enabled remarkable progress in software engineering, particularly in code generation~\cite{chen2021evaluating,austin2021program,chaudhary2023code,gunasekar2023textbooks,muennighoff2023octopack}. However, code editing differs fundamentally from code generation: it requires modifying existing code according to natural language instructions while preserving functionality and minimizing unintended impacts. Despite the successes of LLMs in code generation, recent benchmarks reveal that open-weight and smaller code LMs struggle to perform instruction-guided edits~\cite{muennighoff2023octopack,guo2024codeeditorbench,cassano2024can,singhal2024nofuneval}. This limitation persists even for models pre-trained or fine-tuned on GitHub commit data~\cite{li2023starcoder,lozhkov2024starcoder,muennighoff2023octopack,cassano2024can,xie2025swe}, because commit messages are typically standardized, concise, and lack the linguistic and structural diversity of real-world editing instructions. While commercial models such as GPT-4o achieve stronger performance on editing benchmarks~\cite{cassano2024can,guo2024codeeditorbench,gupta2023grace}, they are often inaccessible for private or enterprise codebases, highlighting \textbf{the need for specialized instruction-tuning pipelines and high-quality, open-source datasets tailored for code editing}.

Current research on code editing has mainly focused on pre-training small models from scratch or fine-tuning on commit data, such as CCT5~\cite{lin2023cct5} and CodeEditor~\cite{li2023codeeditor}, which show limited performance on instruction-guided code editing tasks~\cite{cassano2024can}. A recent trend is to use synthetic data to train models and enhance their performance on specific tasks~\cite{hu2023instructcoder,wei2024selfcodealign,wei2024magicoder}. However, instruction-tuning data specifically designed for code editing remains scarce. Existing synthetic data generation methods, such as OSS-Instruct~\cite{wei2024magicoder}, primarily target code generation and often rely on closed-source commercial LLMs, raising legal, privacy, and reproducibility concerns that limit their applicability in enterprise or domain-specific scenarios. More recent approaches leverage open-source models to generate synthetic data~\cite{wei2024selfcodealign}, but these pipelines still focus on code generation rather than code editing. Importantly, they do not address the unique challenges of code editing, such as producing high-quality (pre-edit code, instruction, post-edit code) triplets with realistic edit styles. Consequently, \textbf{high-quality, open-source instruction-tuning datasets specifically for code editing remain limited, particularly for enterprise or domain-specific applications.}

To address this gap, we propose \mymodel{}, a data synthesis pipeline for instruction tuning of LLMs on code editing tasks. Our approach uses two complementary open-source LLMs to generate pre-edit code and natural-language edit instructions from real code snippets, ensuring diversity and reducing bias. To better reflect developer practice, we synthesize both \emph{lazy} and \emph{descriptive} instruction styles. Outputs from both models are integrated into a unified dataset and refined through a two-stage filtering procedure that removes noise and redundancy. The core novelty of \mymodel{} is twofold: (1) leveraging open-source LLMs to build reproducible, legally safe datasets, and (2) explicitly addressing code editing by generating realistic (pre-edit, instruction, post-edit) triplets with diverse styles and quality control. These innovations establish a general-purpose pipeline for high-quality, open-source code editing datasets supporting both research and industry. Moreover, the synthesis process is adaptable: by replacing the seed code snippets, \mymodel{} can readily target domain-specific or enterprise applications.

Using the described pipeline, we created a lightweight Python instruction-tuning dataset, \mydatasetft{}, designed for code editing. Despite having only 20,000 samples, it significantly boosts model performance with just a few hours of fine-tuning. Unlike datasets from a single model, \mydatasetft{} \textbf{combines data from multiple large models}, and their complementary nature increases task diversity, resulting in a 3\% average gain. The dataset also covers \textbf{a wider range of difficulty levels} and \textbf{features instruction styles closer to real-world editing tasks}. After removing redundant and noisy samples, we found that \textbf{using just one-third of the data yields even better results}, demonstrating a ``less is more'' effect.

We fine-tuned Qwen3-8B-Base, Qwen-2.5-Coder-7B-Base, and DeepSeekCoder-Base-6.7b on \mydatasetft{} and evaluated them on the \canitedit{} benchmark~\cite{cassano2024can}, a rigorous code editing benchmark. Compared to their standard instruction-tuned versions, our models show pass@1 improvements of 4.50\% to 20.79\%, narrowing the gap with closed-source models and falling within 3.54\% of GPT-4---showcasing the effectiveness and competitiveness of our approach.

In summary, our contributions are:
\begin{itemize}
    \item \mymodel{}, a fully open-source data synthesis pipeline for instruction tuning LLMs on code editing via open-source LLMs. 
    \item \mydatasetft{}, a high-quality dataset for code editing, containing 20,000 samples with both \emph{descriptive} and \emph{lazy} instruction styles, enabling efficient fine-tuning.
    \item Comprehensive evaluation and analysis, demonstrating the effectiveness of our synthetic data across multiple LLMs. We explore the impact of fully synthesized code versus commit-based data, multi-LLM integration, instruction styles, and data filtering, providing insights and guidance for future work.
    \item Open-source artifacts, including the dataset, code, and three fine-tuned models (\mymodel{}-Qwen3-8B, \mymodel{}-Qwen2.5-Coder-7B, and \mymodel{}-DeepSeekCoder-6.7B), supporting reproducible research and practical code editing applications.
\end{itemize}

\section{Methodology} 
\label{sec:method}
\begin{figure}
    \centering
    \includegraphics[width=0.8\linewidth]{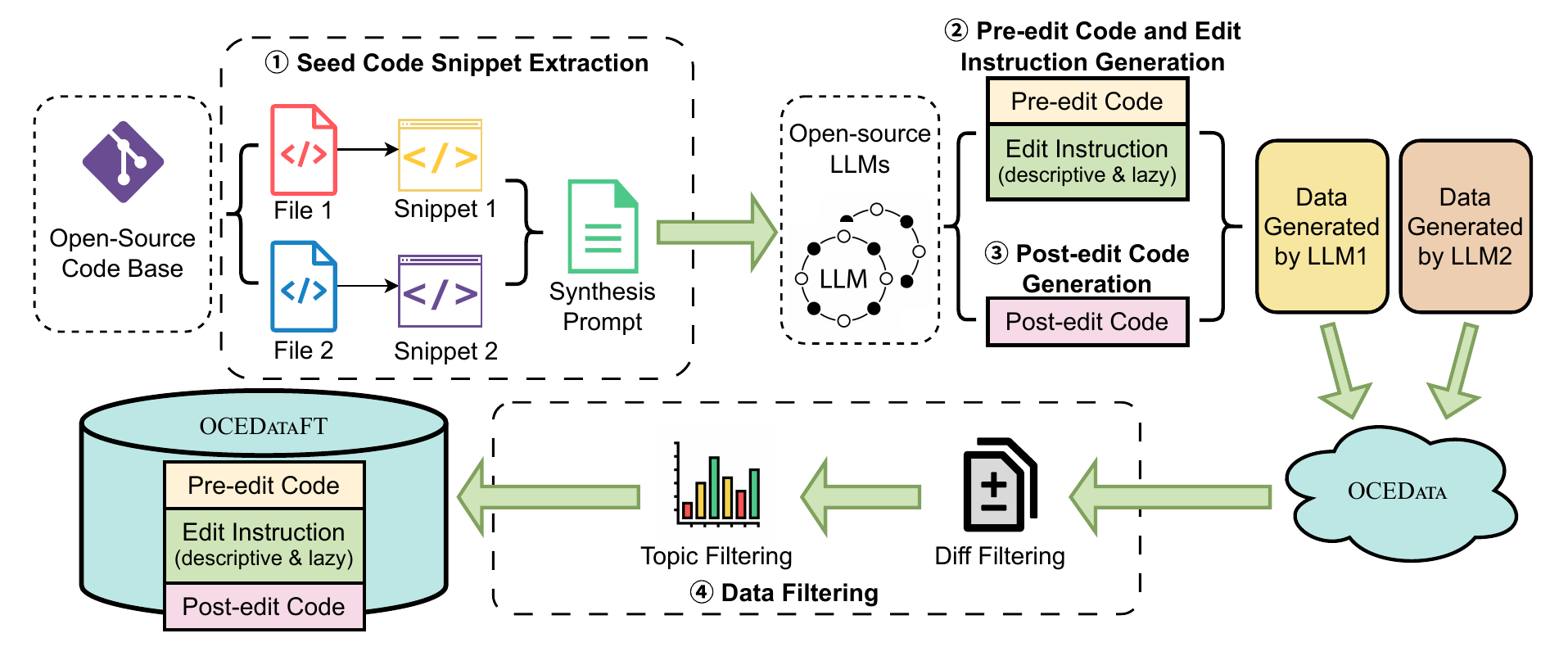}
    \caption{Overview of \mymodel{}.}
    \label{fig:overview_pipeline}
\end{figure}

In this section, we introduce the \mymodel{} pipeline for synthesizing instruction-tuning data for code editing using open-source LLMs. The overall workflow is shown in Figure~\ref{fig:overview_pipeline} and consists of four stages:
\textbf{\ding{172} Seed Code Snippet Extraction}, where authentic code fragments are sampled as the foundation of synthesis;
\textbf{\ding{173} Pre-edit Code and Instruction Generation}, where editable snippets and corresponding natural-language requests are generated;
\textbf{\ding{174} Post-edit Code Generation}, where revised code is produced to fulfill the requested edits; and
\textbf{\ding{175} Data Filtering}, where noisy or redundant samples are removed to ensure dataset quality.
To better reflect real-world editing scenarios, our generated dataset includes both \emph{lazy} and \emph{descriptive} instruction styles, encouraging models to generalize across concise developer prompts and more detailed specifications.

\subsection{Code Editing Dataset Format}
\label{sec:dataset:format}
\begin{figure}
    \centering
    \includegraphics[width=\linewidth]{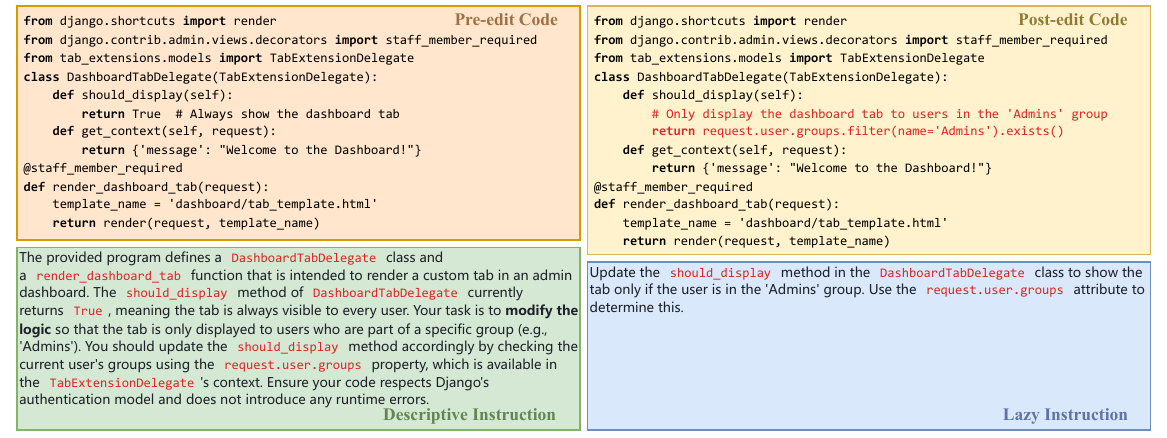}
    \caption{An example of the pre-edit code, post-edit code, and the edit instructions.}
    \label{fig:triplet_example}
\end{figure}

In practical code editing applications, LLMs take as \textbf{input} an original code snippet along with a natural-language edit instruction, and produce as \textbf{output} the revised code snippet that implements the requested change. Figure~\ref{fig:triplet_example} illustrates an example of the code editing training data produced by our pipeline. Each instance is structured as a \textit{code edit triplet}:
\begin{itemize}
\item \textbf{Pre-edit code:} the original snippet requiring modification.
\item \textbf{Edit instruction:} a natural-language description specifying the intended change.
\item \textbf{Post-edit code:} the revised snippet after applying the edit.
\end{itemize}

Formally, each sample is represented as \texttt{(pre-edit code, edit instruction, post-edit code)}, where the first two components serve as model input and the last as the ground-truth output. This triplet directly captures instruction-guided code editing: mapping an existing implementation and a natural-language edit request to the corrected result.

To reflect the diversity of real-world editing scenarios, our dataset includes two complementary instruction styles~\cite{cassano2024can}:
\begin{itemize}
\item \textbf{Lazy instructions}, concise and high-level, resembling developer-written prompts (e.g., ``add error handling for null inputs'').

\item \textbf{Descriptive instructions}, detailed and context-aware, similar to model-generated reflections that fully articulate the required change.
\end{itemize}

Figure~\ref{fig:triplet_example} provides an example of both styles, encouraging models to generalize across terse developer inputs and richer specifications.

\subsection{Seed Code Snippet Extraction}
\label{sec:seed_extract}
The first step in our pipeline is to extract \emph{seed code snippets}, which serve as the foundation for synthesizing realistic and diverse code editing tasks. This step is essential because grounding instruction generation in real code ensures \emph{authenticity} while providing the basis for diverse and non-obvious edit tasks. Our approach is inspired by prior work on instruction synthesis~\cite{wei2024magicoder} and practical applications in code generation, adapted here specifically for code editing scenarios.

To obtain the seed snippets, we randomly select two files from a codebase and extract 5–15 consecutive lines from each, discarding files shorter than 5 lines. Using two snippets in the same prompt allows the LLM to integrate information across different contexts, enhancing the diversity and richness of the generated edits. This design also supports scalability and adaptability: organizations can apply the same method to their internal codebases to create instruction-tuning datasets aligned with proprietary coding styles and domain-specific requirements, enabling realistic, enterprise-focused code editing tasks.

\subsection{Pre-edit Code and Edit Instruction Generation}
\label{sec:pre-edit-gen}
\begin{figure}[htbp]
\centering
\includegraphics[width=1.0\linewidth]{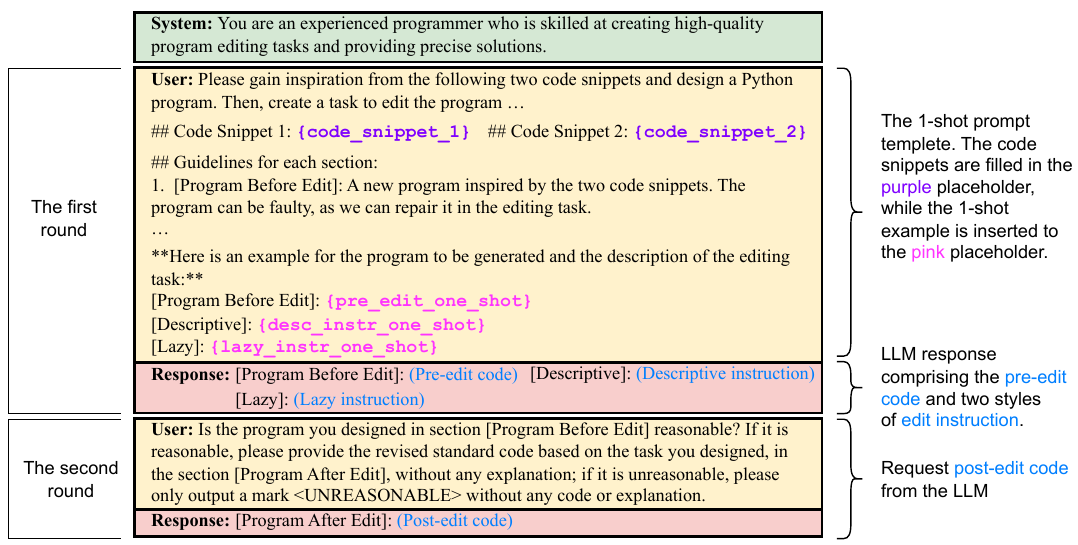}
\caption{An example of the two-round dialogue.}
\label{fig:chat_example}
\end{figure}

The first round of our two-round dialogue focuses on generating both the \textbf{pre-edit code} and the corresponding \textbf{edit instructions}, laying the foundation for realistic instruction-tuning data while maintaining privacy and diversity. An example of this dialogue is shown in Figure~\ref{fig:chat_example}.

We employ two distinct open-source LLMs as instruction generators. Since these models have been pre-trained on different code corpora, they bring complementary strengths. Using both models increases diversity, reduces model-specific biases, and avoids reliance on closed-source commercial LLMs, thereby mitigating legal, privacy, and reproducibility concerns.

The generation is guided by a carefully designed 1-shot prompt template that incorporates the synthesis instruction, the two seed code snippets extracted in Section~\ref{sec:seed_extract}, and a randomly selected 1-shot example from a curated pool of 20 instances. The seed snippets inspire the pre-edit code, ensuring contextually meaningful tasks while minimizing the risk of sensitive information leakage. Randomly selecting the 1-shot example further enhances output diversity and prevents overfitting to a fixed demonstration.

To better capture the diversity of real-world editing scenarios, we generate both \emph{lazy} and \emph{descriptive} instructions in a single generation pass. Producing both styles simultaneously improves efficiency, ensures consistent alignment with the same pre-edit code, and provides complementary perspectives that help models generalize across terse developer prompts and more detailed, context-aware specifications.

Figure~\ref{fig:triplet_example} illustrates a generated task about administration control derived from two seed snippets: one implementing an admin extension with tab display logic, the other utilizing a static method decorator for markup rendering, which is quite different. The resulting pre-edit code and paired lazy and descriptive instructions together form a coherent editing task suitable for subsequent post-edit code generation. Note that two LLMs could generate quite different pre-edit code and edit instruct even given the same seed snippets, for example, another model presents a program that defines a Django admin extension with a tab delegate and a placeholder renderer class, with an edit task requires fixing a quote mismatch, adding a render method to the renderer class, and implementing a tab content method in the delegate class.

\subsection{Post-edit Code Generation}
In the second round of dialogue, we generate the \textbf{post-edit code} based on the pre-edit code and edit instruction produced in the first round (Section~\ref{sec:pre-edit-gen}). The LLMs are prompted to produce a revised code snippet that fulfills the specified editing task, ensuring consistency with the provided instruction and pre-edit context.

To enhance reliability, we introduce a self-checking mechanism: the LLM evaluates whether the pre-edit code and edit instruction constitute a reasonable editing task. If the task is deemed ill-posed, the model outputs a special token \texttt{<UNREASONABLE>}. Otherwise, it generates the post-edit code.

Outputs from two open-source LLMs are then merged to form the mixed dataset \mydataset{}, improving linguistic and stylistic diversity. Using two models leverages complementary pretraining knowledge, reduces systematic biases, and enhances overall data quality. This two-round dialogue process produces high-quality (pre-edit code, edit instruction, post-edit code) triplets suitable for instruction-tuning, while ensuring that the post-edit code is generated afresh and guided entirely by the instruction, not by copying from the seed snippets.

\subsection{Data Filtering}
\label{sec:data_filtering}
High-quality data is essential for effective instruction tuning of LLMs~\cite{zhou2023lima,wang2024codeclm,kopf2023openassistant,chen2024alpagasus}, whereas noisy or redundant samples increase training cost and degrade model performance. Because synthetic code generated by LLMs often suffers from such issues, we introduce \myfiltering{}, a two-step procedure consisting of \textbf{diff filtering} and \textbf{topic filtering}, to systematically improve the quality of code editing datasets.

\subsubsection{Diff Filtering}
To identify noisy or overly complex samples, we analyze the differences between pre-edit and post-edit code using the \texttt{difflib} Python library~\cite{python-difflib}. This library provides efficient sequence-matching algorithms that compute differences between two text sequences, widely used for generating unified diffs in version control systems. Leveraging this functionality, we extract two measures of edit complexity: (1) the number of modified lines, including additions, deletions, and revisions, and (2) the number of hunks, where a hunk is defined as a contiguous block of changes together with three surrounding context lines. The analysis is conducted on the mixed dataset generated by two models (see Section~\ref{sec:implement} for implementation details).

Figure~\ref{fig:modified_lines_distrib} and Figure~\ref{fig:hunk_distrib} present the distribution of edit complexity. Most synthesized edits involve 5–20 modified lines and 1–2 hunks, reflecting focused modifications. However, the distribution is long-tailed: a small fraction exceeds 70 modified lines or contains more than 7 hunks, representing highly dispersed and challenging tasks. Such instances are less suitable for instruction tuning, as overly complex edits may hinder model learning~\cite{shen2024efficient,hu2021model,mina2025cognitive}.

Based on this analysis, we discard samples with more than 70 modified lines or more than 7 hunks, as well as samples with zero hunks (i.e., post-edit code identical to pre-edit code). This filtering step ensures that retained data represent meaningful and learnable code edits.

\subsubsection{Topic Filtering}
To enhance diversity and reduce redundancy, we perform topic analysis on the concatenation of pre-edit code and edit instructions. For this purpose, we adopt Hierarchical Dirichlet Process (HDP) modeling~\cite{iammarino2020topic,silva2021topic,carissimi2025towards}, a nonparametric Bayesian method that automatically infers the number of latent topics without requiring pre-specification.

\begin{figure}[htbp]
  \centering
  \begin{subfigure}[t]{0.32\linewidth}
    \centering
    \includegraphics[width=\linewidth]{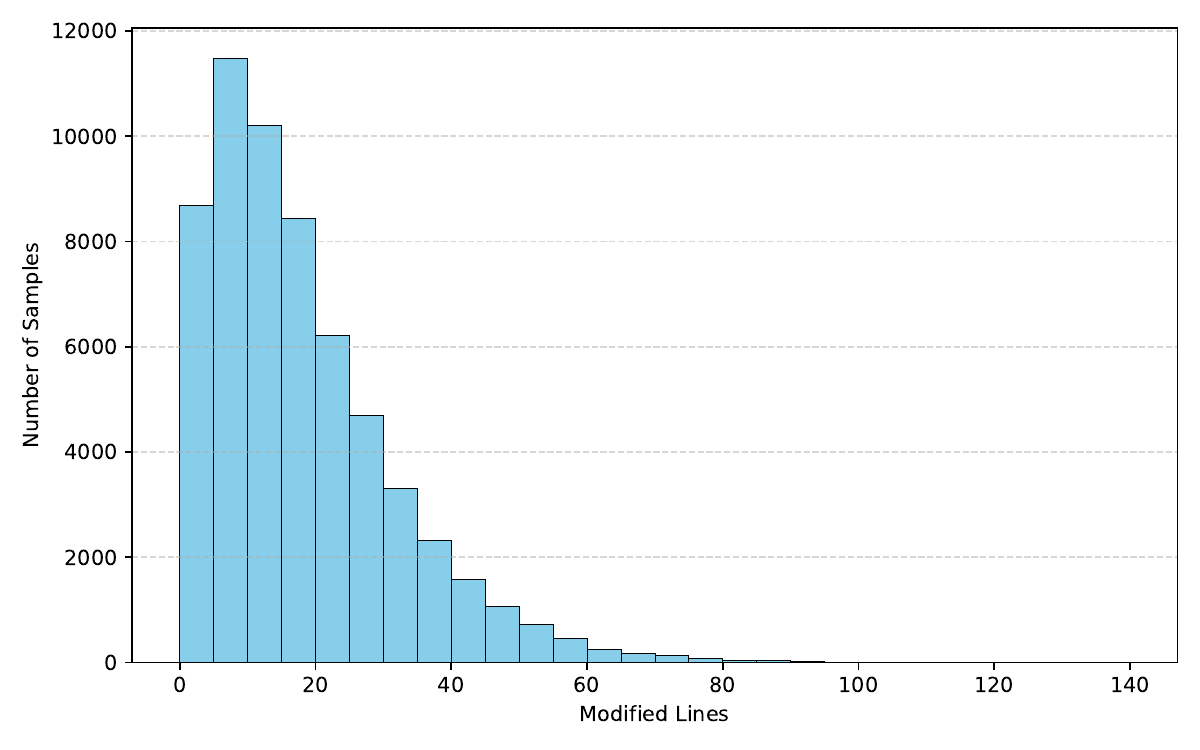}
    \caption{Modified lines}
    \label{fig:modified_lines_distrib}
  \end{subfigure}
  \hfill
  \begin{subfigure}[t]{0.32\linewidth}
    \centering
    \includegraphics[width=\linewidth]{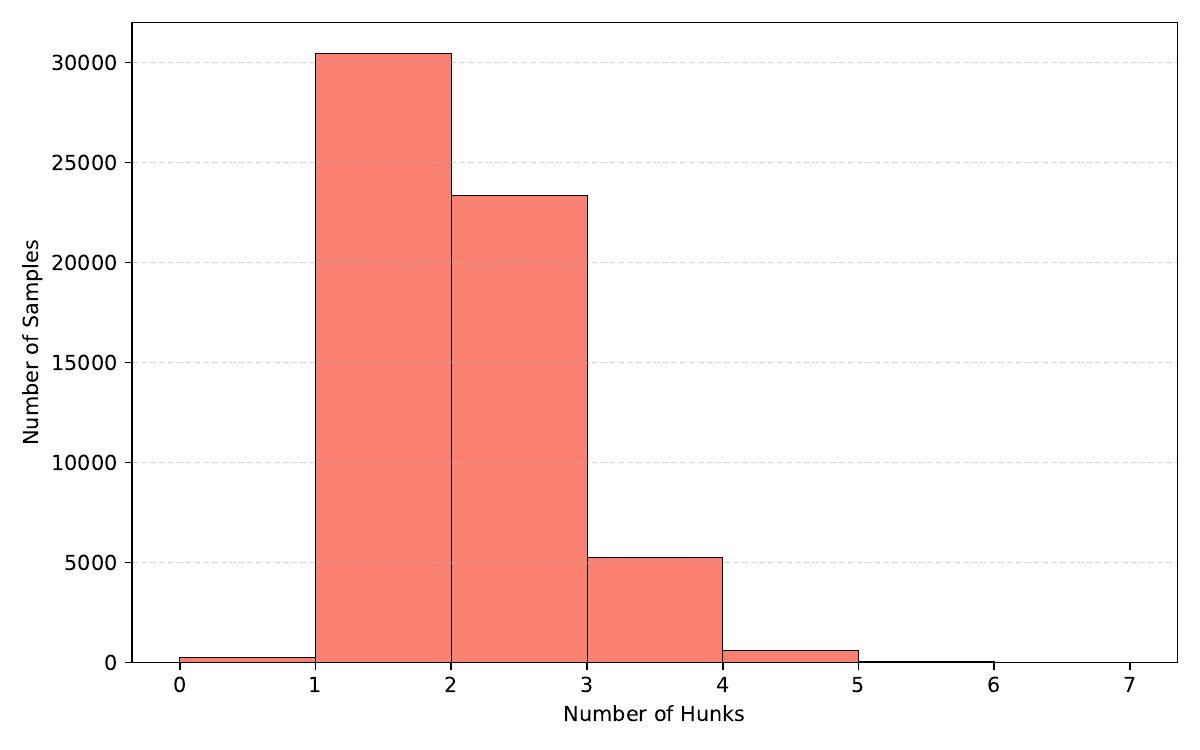}
    \caption{Number of hunks}
    \label{fig:hunk_distrib}
  \end{subfigure}
  \hfill
  \begin{subfigure}[t]{0.32\linewidth}
    \centering
    \includegraphics[width=\linewidth]{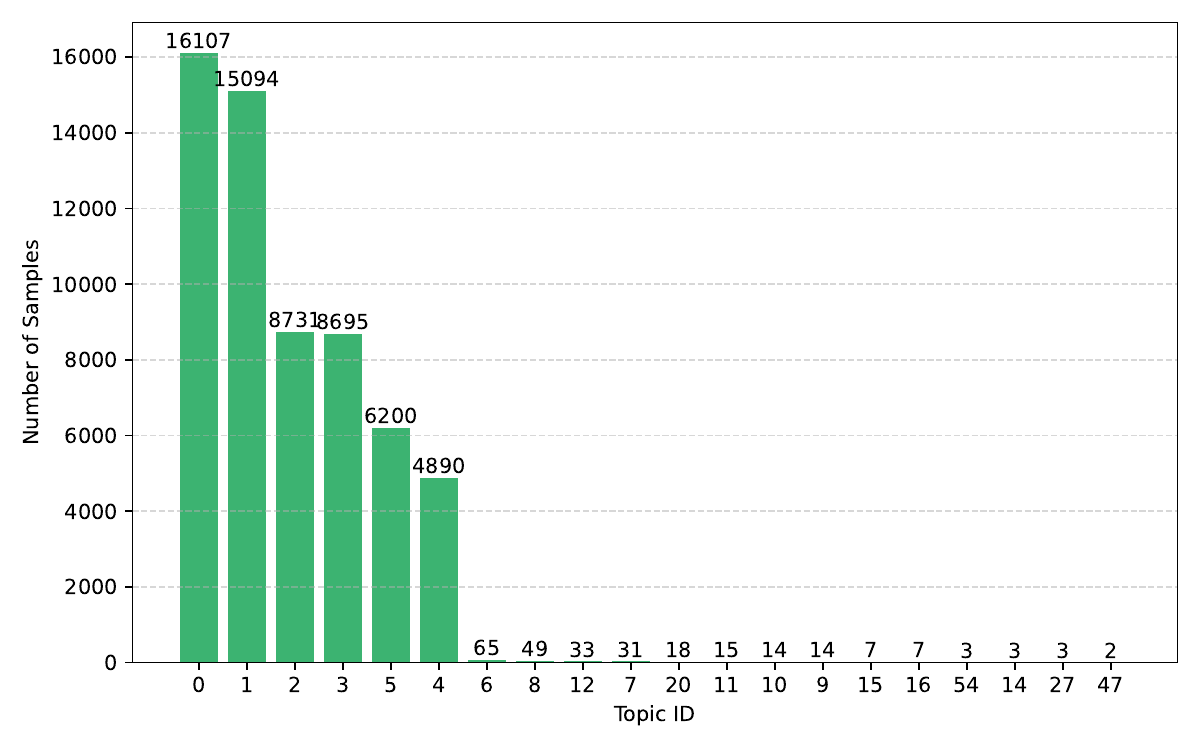}
    \caption{HDP topic distribution}
    \label{fig:topic_distrib}
  \end{subfigure}
  \caption{Distribution of modified lines, hunks, and topics in the combined dataset from Qwen3 and DeepSeek.}
  \label{fig:data_distrib}
\end{figure}

Our analysis of LLM-generated outputs shows that a substantial portion of instances cluster into only a few dominant topics (Figure~\ref{fig:topic_distrib}), which indicates limited topical variety and redundancy across samples. Since topic diversity is critical for instruction tuning, and models trained on narrowly distributed topics may struggle to generalize to unseen editing scenarios, we introduce a quota-based allocation strategy. This method selectively reduces the number of samples from overrepresented topics while preserving those from underrepresented ones according to the target number of samples. 
For example, assuming that we have 50 samples in four topics: A (25), B (15), C (7), and D (3), and want to reduce the total to 20 while keeping all topics represented. In each round, a per-topic quota is assigned; topics with fewer samples than their quota are locked and kept entirely, while the remaining quota is redistributed among the other topics. After repeating this process, the final counts are A (6), B (6), C (5), and D (3), ensuring smaller topics are not underrepresented. The full algorithm is available in our repository.
In this way, the final dataset achieves a more balanced topical distribution, better coverage of diverse editing patterns, and reduced risk of overfitting to repetitive tasks.

\subsubsection{Summary of \myfiltering{}}
Based on the above analysis, the filtering procedure is as follows:
\begin{enumerate}
\item Remove instances with more than 70 modified lines or more than 7 hunks to control complexity;
\item Remove instances with zero hunks to eliminate noisy data;
\item Filter samples from overrepresented topics according to the target number (e.g., 20,000) to ensure balance and diversity.
\end{enumerate}

This integrated approach guarantees that the final dataset consists of high-quality, diverse, and realistically structured code edit triplets suitable for instruction tuning.

\section{Implementation Details}
\label{sec:implement}
This section presents the implementation details of our experiments. 

\subsection{Seed Code Snippet Extraction}
We adopt \commitpack{} \cite{muennighoff2023octopack} as our seed corpus in this work. \commitpackft{} is filtered from \commitpack{}, a code instruction dataset constructed by the commits scraped from GitHub. We choose \commitpackft{} because the dataset is collected from the open-source community and has been filtered by a series of rules, ensuring it is checked without the risk of data leakage. However, we only extract code snippets from the code before commit, rather than using the entire commit, due to the relatively low quality of the commit data (analyses are shown in Section~\ref{sec:gen_code}). In our study, we use the Python subset of \commitpackft{}.

\subsection{Data Synthesis}
\label{sec:data_synthesis}
In constructing the dataset, we employ the Qwen3-32B-Instruct and DeepSeek-V3-0324 models to generate each part of the code edit triplet, separately. The temperature is set to 0.8 and the top-p value to 0.95, with a maximum number of output tokens of 2048. For each pair of code snippets, we generate data only once. Due to hardware constraints, we conduct LLM inference through API calls from DeepSeek~\cite{deepseek_platform_2024} and Aliyun~\cite{alibaba_cloud_2024}; however, deploying these models locally for inference also constitutes a viable alternative. 

We generate 30,000 samples from each LLM, each including both a descriptive and a lazy edit instruction, yielding 60,000 code edit triplets of the form \texttt{(pre-edit code, edit instruction, post-edit code)} for DeepSeek-V3 and Qwen3-Instruct, referred to as \mydataset{}-DS and \mydataset{}-Qwen3, respectively. From each dataset, we randomly select 15,000 descriptive and 15,000 lazy samples, combining them into a mixed dataset, \mydataset{}, which comprises a total of 60,000 samples.

\subsection{Data Filtering}
For diff filtering, we use the \texttt{SequenceMatcher} from the \texttt{difflib} Python library to compare pre-edit and post-edit code, and then calculate the number of modified lines and hunks. For topic filtering, we use the \texttt{HdpModel} from the \texttt{gensim} Python library. Stopwords for natural language are obtained from the \texttt{nltk} library. Since no standard list of stopwords exists for code, we constructed one manually by collecting common reserved words across programming languages.

\subsection{Model Training}
\label{sec:model_training}
\begin{figure}
    \centering
    \includegraphics[width=0.8\linewidth]{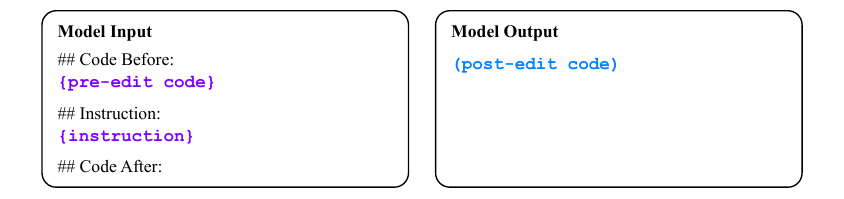}
    \caption{The input-output format of model training and evaluation.}
    \label{fig:finetune_prompt}
\end{figure}

In our experiments, we train multiple models using a variety of datasets, including our \mydatasetft{}.
We convert each synthesized edit example into a supervised instruction-tuning pair for fine-tuning. For each edit triplet \texttt{(pre-edit code, edit instruction, post-edit code)}, we serialize it into a compact input prompt and a corresponding ground-truth output for model training. Following~\cite{lozhkov2024starcoder,cassano2024can}, the input-output format of the code edit task is shown in Figure~\ref{fig:finetune_prompt}. We then fine-tune the base models using LLaMA-Factory~\cite{zheng2024llamafactory} with LoRA (rank = 8). Training is performed on an A800 GPU for 2 epochs, with a batch size of 1 per GPU, a sequence length of 2048, and the Adam optimizer (learning rate $1\times10^{-4}$). A cosine scheduler with a 0.1 warmup ratio is applied. 

\section{Evaluations}
\label{sec:eval}
In this section, we comprehensively evaluate \mymodel{} by answering 6 research questions.

\begin{itemize}
    \item \textbf{RQ1 (Effectiveness)}: How effectively can \mymodel{} generate a fine-tuning dataset (\mydatasetft{}) that improves code editing across multiple LLMs?
    \item \textbf{RQ2 (Synthesizing Data Versus Original Commit)}: How does generating entirely new pre-edit and post-edit code compare to using original commit data in improving model performance?
    \item \textbf{RQ3 (Multi-LLM Data Integration)}: To what extent does integrating data from multiple LLMs improve fine-tuned models?
    \item \textbf{RQ4 (Integration of Descriptive and Lazy Instruction Styles)}: How does combining \emph{descriptive} and \emph{lazy} instruction styles affect model generalization compared to using a single style?
    \item \textbf{RQ5 (Impact of \myfiltering{} on Data Quality Enhancement)}: Can the \myfiltering{} method enhance data quality and improve fine-tuned model performance?
\end{itemize}

\subsection{Benchmark and Evaluation Settings}
\label{sec:benchmark_eval_settings}
\canitedit{}~\cite{cassano2024can} is a benchmark designed to evaluate the code editing capabilities of LLMs. It comprises 105 manually curated Python problems, each accompanied by two types of instructions: a concise lazy edit instruction and a detailed descriptive edit instruction. The tasks are evenly distributed across bug fixes, feature enhancements, and new feature additions, covering domains such as algorithms, data processing, and game programming, and involving libraries such as NumPy and PyTorch. Each problem is paired with hidden test cases for correctness verification.

Following previous work~\cite{muennighoff2023octopack,lozhkov2024starcoder,hui2024qwen2}, we evaluate model performance using the \emph{pass@1} metric~\cite{chen2021evaluating}, which measures the probability that a single generated solution passes all predefined test cases. Formally, for a set of model-generated answers, pass@1 is defined as:
\[ \mathrm{pass@1} = \frac{1}{N}\sum_{i=1}^{N} \mathbf{1}\left [ \mathrm{Solution}  \; i \; \mathrm{passes \; all \; tests}  \right ]  \]
where $N$ denotes the number of generated solutions and $\mathbf{1}[\cdot]$ is the indicator function. A higher pass@1 value indicates a greater success rate, thereby reflecting stronger instruction-following and code editing performance.

We use the settings following~\cite{cassano2024can}: 2048 maximum new tokens, temperature of 0.2, and top-p of 0.95. We sample 20 completions for each problem, and calculate pass@1. The input-output format in evaluation is identical to model training, as shown in Figure~\ref{fig:finetune_prompt}.

\subsection{Choice of Post-Filtering Data Scale for \mydatasetft{}}
\label{sec:choice_data_scale}

To identify the optimal data size for \myfiltering{}, we fine-tune Qwen3-8B-Base on filtered datasets of varying scales. Filtering is first applied separately to descriptive and lazy instructions in \mydataset{}, yielding 29,716 and 29,697 instances, respectively. The HDP filtering process then removes redundant data from each subset according to the target size, and the two subsets are merged to maintain a 1:1 ratio.

As shown in Table~\ref{tab:different_train_data}, fine-tuning with 20,000 samples achieves the best Pass@1 performance on the \canitedit{} dataset, outperforming both the unfiltered 60,000-sample setting and the 30,000-sample setting. In contrast, using only 10,000 samples leads to underfitting. These results suggest that redundant data harms model performance, while insufficient data limits learning. Accordingly, we use 20,000 instances for fine-tuning in \mydatasetft{} and in most subsequent experiments.

\begin{table}[htbp]
    \caption{Pass@1 (\%) of model fine-tuned with different amount of data on \canitedit{} benchmark.}
    \label{tab:different_train_data}
    \centering
    \scriptsize
    \begin{tabular}{llccc}
    \toprule
     Base Model & Data Amount & Lazy&Descriptive  &Overall
\\
    \midrule
    \multirow{5}{*}{Qwen3-8B-Base} & 5,000 & 45.90 &57.38 &51.64 
\\
     & 10,000 & 46.29 &58.95 &52.62 
\\
     & 30,000 & 47.43 &58.71 &53.07 
\\
     & 60,000 (unfiltered) & 45.00 & 58.52 &51.76 
\\
 & 20,000 & \textbf{48.14} &\textbf{60.05} &\textbf{54.10} 
\\
    \bottomrule
    \end{tabular}
\end{table}

\subsection{RQ1: Effectiveness}
To evaluate the effectiveness of \mymodel{}, we benchmark the models fine-tuned on \mydatasetft{} against several state-of-the-art LLMs.
\subsubsection{Design}
\textbf{Base Models. }
We fine-tune three state-of-the-art base models, namely Qwen3-8B-Base~\cite{yang2025qwen3}, Qwen-2.5-Coder-7B-Base~\cite{hui2024qwen2}, and DeepSeekCoder-Base-6.7B~\cite{guo2024deepseek}, on our \mydatasetft{} dataset to evaluate the effectiveness of our data generation pipeline. The fine-tuned models are denoted as \mymodel{}-Qwen3-8B, \mymodel{}-Qwen2.5-7B, and \mymodel{}-DSC-6.7B, respectively (collectively referred to as \mymodel{}-series). Following Section~\ref{sec:choice_data_scale}, we use 20,000 samples for fine-tuning, with parameter settings in Section~\ref{sec:model_training}.

\textbf{Baseline Selection. }
For a fair and comprehensive evaluation, we compare our \mymodel{}-series models with a diverse set of baseline methods. Our selection encompasses a diverse range of models, including both open-source and proprietary (closed-source) offerings, as well as instruction-tuned models designed for various downstream tasks. Specifically, we include:
\begin{itemize}
    \item Open-sourced code LLMs for general code tasks: CodeLlama-Instruct-7B ~\cite{roziere2023code}, SelfCodeAlign-CQ-7B~\cite{wei2024selfcodealign}, DeepSeekCoder-Instr-6.7B~\cite{guo2024deepseek}, and Qwen-2.5-Coder-7B-Instr~\cite{hui2024qwen2}.
    \item Open-sourced general LLM: Qwen3-8B-Instr~\cite{yang2025qwen3}.
    \item Closed-sourced LLM: GPT-4~\cite{openai2023gpt4} and GPT-3.5-Turbo~\cite{openai2023chatgptapi}.
    \item Open-sourced code LLM for code editing task: Editcoder-6.7B and -33B\cite{cassano2024can}.
\end{itemize}

To the best of our knowledge, Editcoder is the only open-sourced LLM specifically designed for code editing tasks. Therefore, we selected Editcoder with two parameter sizes along with several code LLMs for general code tasks as baselines. 
Qwen3-8B-Instr is a SOTA general-purpose instruction-tuned model, which is fine-tuned from Qwen3-8B-Base. The latter serves as a SOTA base model and will also be used as the base model for instruction-tuning in subsequent experiments.
To enable a more comprehensive performance comparison, we also included the closed-source GPT-4 and GPT-3.5-Turbo from OpenAI as a baseline model. These models are benchmarked using their APIs~\cite{openai-api}.

For all the models, we benchmark them on \canitedit{} using the settings in Section~\ref{sec:benchmark_eval_settings}.

\subsubsection{Results}
Based on the comprehensive results in Table~\ref{tab:effectiveness}, several conclusions can be drawn.

\begin{table}[htbp]
    \caption{Pass@1 (\%) of the fine-tuned models on \canitedit{} benchmark.}
    \label{tab:effectiveness}
    \centering
    \scriptsize
    \begin{tabular}{lllccc}
    \toprule
    Base Model & Model Name & Dataset & Lazy& Descriptive &Overall
\\
    \midrule
    --           & GPT-4         & Proprietary & \textbf{51.95} & \textbf{63.33} & \textbf{57.64} \\
    --           & GPT-3.5-Turbo & Proprietary & 42.71 & 48.14 & 45.43 \\
    \midrule
    DeepSeekCoder-Base-33B & DeepSeekCoder-Instr-33B & Not disclosed & 42.33 & 55.90  &49.12 \\
    \midrule
    CodeLlama-7B & CodeLlama-Instruct-7B & Proprietary & 23.49 & 32.83  &28.16 \\
    \midrule
    CodeQwen1.5-7B & SelfCodeAlign-CQ-7B & \makecell[l]{Generated by\\ SelfCodeAlign} & \makecell[c]{Not\\Provided} & \makecell[c]{Not\\Provided}  &39.00 \\
    \midrule
    \multirow{4}{*}{DeepSeekCoder-Base-6.7B} & DeepSeekCoder-Instr-6.7B & Not disclosed & 31.65 & 41.03  &36.34 
\\
     & Editcoder-6.7B & \makecell[l]{EditPackFT, \\ Commit2023FT} & 39.29 & 48.33  &43.81 
\\
     & \mymodel{}-DSC-6.7B & \mydatasetft{} & \textbf{43.90} & \textbf{52.81}  &\textbf{48.36} 
\\
    \midrule
    \multirow{2}{*}{Qwen2.5-Coder-7B-Base} & Qwen-2.5-Coder-7B-Instr & Not disclosed & 43.00 & 51.43  &47.21 
\\
     & \mymodel{}-Qwen2.5-7B & \mydatasetft{} & \textbf{47.19} & \textbf{56.24}  &\textbf{51.71} \\
     \midrule
    \multirow{2}{*}{Qwen3-8B-Base} & Qwen3-8B-Instr & Not disclosed & 32.81 & 33.81  & 33.31 
\\
     & \mymodel{}-Qwen3-8B & \mydatasetft{} & \textbf{48.14} & \textbf{60.05}  &\textbf{54.10} 
\\
    \bottomrule
    \multicolumn{6}{l}{\makecell[l]{\textit{Note: }The results of GPT-4, GPT-3.5, DeepSeekCoder-Instr-33B, DeepSeekCoder-Instr-6.7B, CodeLlama-Instruct-7B, \\
    and Editcoder-6.7B are cited from~\cite{cassano2024can}; 
    the result of SelfCodeAlign-CQ-7B is cited from~\cite{wei2024selfcodealign}}}
    \end{tabular}
\end{table}

\textbf{Cross-Architecture Generalizability.}
\mymodel{} is model-agnostic, providing a unified pipeline to generate instruction tuning data (\mydatasetft{}) tailored for code editing tasks. By leveraging smaller but higher-quality datasets specifically designed for editing, it achieves consistent and substantial improvements across architectures: \mymodel{}-series outperform the corresponding instruction models in pass@1, ranging from 4.50\% to 20.79\%. This stands in contrast to many existing instruction-tuned models, whose large-scale training data are undisclosed and mainly general-purpose, making them less effective for specialized editing scenarios. Moreover, our data generation strategy delivers strong performance gains across different model families and scales, highlighting its broad applicability for enhancing code editing capabilities.

\textbf{Superiority over General Code Models.}
The results in Table~\ref{tab:effectiveness} confirm that \mymodel{}-series models consistently outperform general instruction-tuned models across diverse architectures. The significant gains---ranging from 4.5 to over 10 percentage points---highlight that overlooking fine-tuning on code editing tasks can severely constrain performance, underscoring the importance of this often neglected task. Notably, \mymodel{}-DSC-6.7B achieves 48.36\%, approaching the performance of the much larger DeepSeekCoder-Instr-33B (49.12\%) despite having significantly fewer parameters, which further underscores the efficiency of our approach. Importantly, on the lazy style editing tasks, \mymodel{}-DSC-6.7B achieves 43.90\%, surpassing DeepSeekCoder-Instr-33B (42.33\%), demonstrating superior performance in this specific editing paradigm.

\mymodel{}-Qwen2.5-7B (51.71\%) also exceeds SelfCodeAlign-CQ-7B (39.00\%), despite both using synthetic data. The key difference is that \mydatasetft{} attaches importance to the code editing task, while SelfCodeAlign neglects this task.

\textbf{Beyond Natural Code Change Learning.}
Beyond general models for code tasks, \mymodel{}-series model also surpasses specialized LLM designed for code editing. On the DeepSeek-Coder-6.7B base, \mymodel{}-DSC-6.7B (48.36\%) outperforms Editcoder-6.7B (43.81\%), which is fine-tuned specifically on curated code change datasets (EditPackFT and Commit2023FT). EditPackFT and Commit2023FT are two commit datasets similar to \commitpackft{}, with a total amount of 46,274 instances~\cite{cassano2024can}. Due to the inherently low quality of commit data and their stylistic inconsistencies with actual edit instructions in real-world code editing tasks, using natural commit data directly fails to yield satisfactory results---these factors will be analyzed in detail in RQ2.

\textbf{Near-Parity with GPT-4 in Code Editing.}
\mymodel{} demonstrates strong performance on newer model architectures, as shown by \mymodel{}-Qwen3-8B, which achieves 54.10\% on \canitedit{}, significantly outperforming its instruction-tuned counterpart Qwen3-8B-Instr (33.31\%). This substantial gain confirms the effectiveness of our method even on advanced, non-specialized base models. Moreover, \mymodel{}-Qwen3-8B approaches the performance of closed-source models---its score is close to that of GPT-4---demonstrating that our approach can achieve competitive, near-state-of-the-art results on code editing tasks without relying on proprietary models or massive-scale training.

\finding{\mymodel{} achieves strong and consistent gains across model families and scales by focusing on realistic code editing tasks. It outperforms general instruction-tuned models and specialized code editors through synthetic data derived from code snippets, demonstrating that task-aligned data generation is key to effective code editing.}

\subsection{RQ2: Synthesizing Data Versus Original Commit}
\label{sec:gen_code}
To demonstrate the necessity of generating entirely new pre-edits, post-edits, and edit instructions, we fine-tune Qwen3-8B-Base using datasets under varying configurations.

\subsubsection{Design}
Given the inherent alignment between Git commit structures and code editing tasks, we investigate how substituting edit instructions and code changes with either original commit content or LLM-generated alternatives affects model performance. To this end, we fine-tune Qwen3-8B-Base, which serves as the best-performing base model for RQ1, on 20,000 instances under four data settings that systematically vary the source of edit instructions and code changes, as detailed below:
\begin{itemize}
    \item \textbf{Original Commit:} To establish a baseline performance using raw commit data, all components---pre-edit code, edit instruction, and post-edit code---are directly extracted from the original commit without any modification.
    \item \textbf{Rewritten Commit:} To assess whether reformulating commit messages into natural, actionable instructions improves fine-tuning effectiveness, the edit instruction is rewritten into user-facing formats (descriptive and lazy styles) by Qwen3-32B-Instruct while the pre-edit and post-edit code are retained from the original commit. The pre-edit code, post-edit code, and the original commit message are fed to Qwen3-32B-Instruct for instruction reformulation.
    \item \textbf{Generated Task:} To investigate whether LLM-synthesized instructions and edits can enhance model learning when grounded in real pre-edit contexts, only the pre-edit code is taken from the original commit, while both the edit instruction and post-edit code are generated by Qwen3-32B-Instruct.
    \item \textbf{Generated using \mymodel{} (\mydataset{}):} To evaluate the effectiveness of our proposed approach---which leverages context from multiple code snippets to generate more diverse code edit triplets---we employ the \mymodel{} pipeline. In this pipeline, an LLM generates both the pre-edit code and the edit instruction from a combination of two original commit snippets, followed by the generation of the corresponding post-edit code.
\end{itemize}

To ensure fair comparisons and minimize confounding factors, no filtering procedures were applied in any of the aforementioned data settings.

\subsubsection{Results}
The evaluation results are presented in Table~\ref{tab:gen_code}. 
Overall, the dataset generated with all three components of the code edit triplets (\mydataset{}) achieves the best performance. Unexpectedly, fine-tuning on raw commit data yields even lower performance than the base model, as the noisy and low-quality commits hinder model learning. Simply rewriting commit messages into the form of edit instructions does not improve fine-tuning effectiveness. In contrast, regenerating both the edit instruction and the post-edit code leads to a clear improvement in pass@1, even when the pre-edit code is drawn from the original commit.

\begin{figure}[htbp]
  \centering
  \begin{subfigure}[t]{0.47\linewidth}
    \centering
    \includegraphics[width=\linewidth]{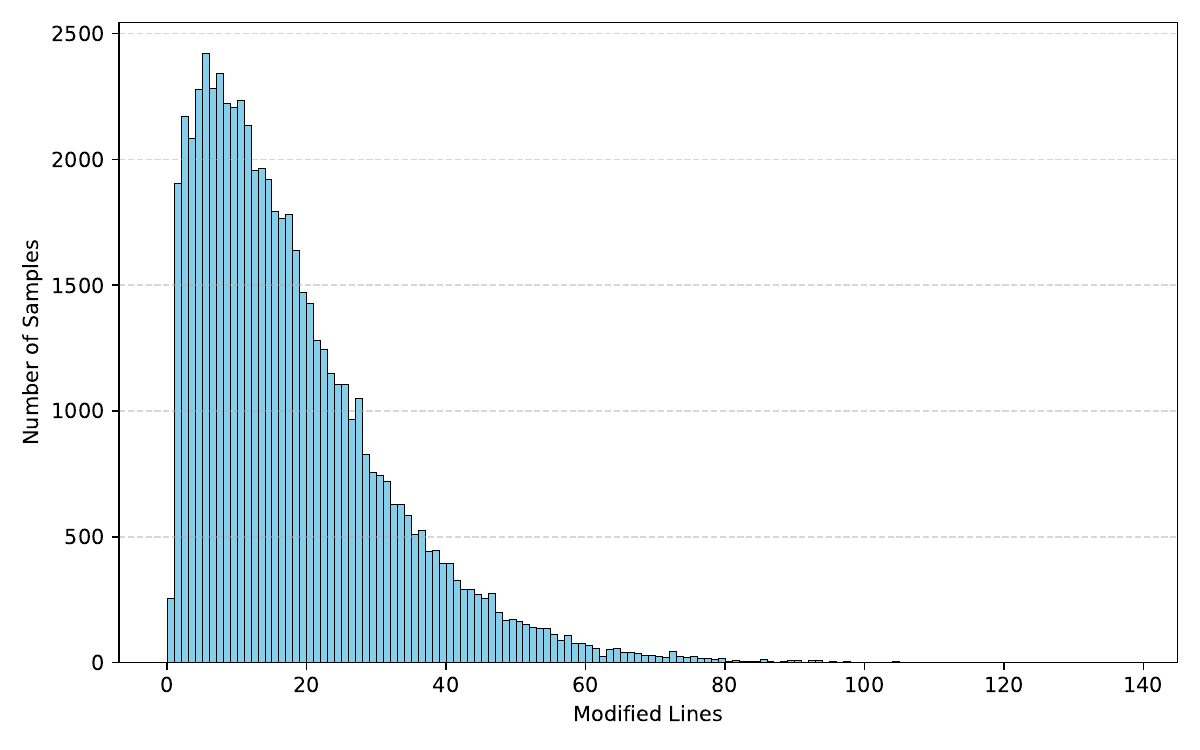}
    \caption{Generated using \mymodel{}}
  \end{subfigure}
  \hfill
  \begin{subfigure}[t]{0.47\linewidth}
    \centering
    \includegraphics[width=\linewidth]{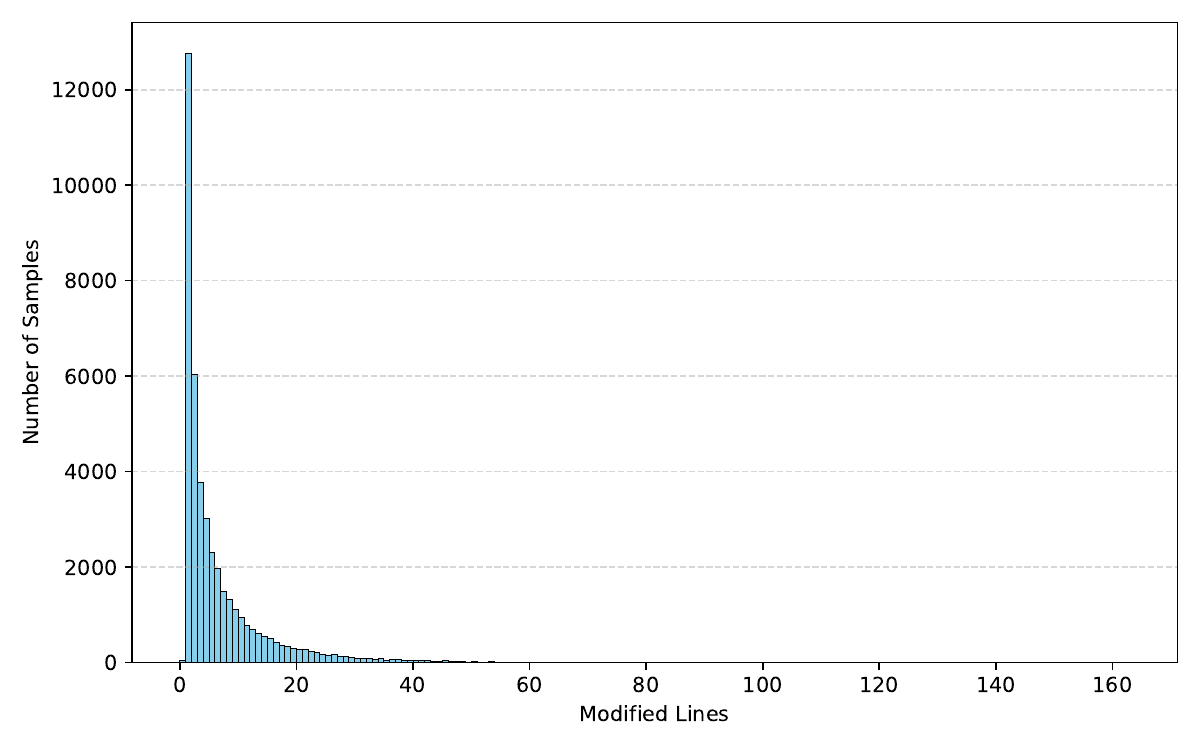}
    \caption{commit from \commitpackft{}}
  \end{subfigure}
  \caption{Distributions of modified lines of commit and data generated using \mymodel{}. For the commit data from \commitpackft{}, we remove those who have empty pre-edit code. }
  \label{fig:modified_lines_commit}
\end{figure}

Based on the assumption that original commits mostly contain simple tasks, we examine the \diff{} between pre-edit and post-edit code for all commits compared with \mydataset{} (Figure~\ref{fig:modified_lines_commit}), excluding commit-only data. We find that the vast majority of commits modify just a single line, resulting in overly simplistic editing tasks. In real-world scenarios, however, code edits typically span multiple lines. Merely converting commit messages into edit instructions does little to address this simplicity and may even introduce additional noise. By contrast, \mydataset{} exhibits a more balanced distribution of modification sizes, with most edits ranging from 1 to 20 lines, thereby producing tasks that better reflect real-world complexity.

\begin{table}[htbp]
    \caption{Pass@1 (\%) of models fine-tuned on 20,000 data points across various datasets}
    \label{tab:gen_code}
    \centering
    \scriptsize
    \begin{tabular}{lllllccc}
    \toprule
     Base Model &  Data Setting&Pre-edit Code& Post-edit Code &Edit Instruction& Lazy& Descriptive &Overall
\\
    \midrule
        \multirow{5}{*}{Qwen3-8B-Base} & Without Finetune & -- & -- & -- & 38.14 & 50.33
 & 44.24
\\
     &   Original Commit&Commit& Commit&Commit& 34.95 & 45.57 
 &40.26 
\\
     &  Rewritten Commit&Commit& Commit& Rewritten& 32.19 &45.33 
 &38.76 
\\
     &   Generated Task&Commit& Generated&Generated& 41.24 & 48.95
 &45.10 
\\
     & \mydataset{} (unfiltered) &Generated& Generated& Generated& \textbf{44.38} & \textbf{54.43}
 &\textbf{49.40} 
\\
    \bottomrule
    \end{tabular}
\end{table}

\finding{LLMs fine-tuned on synthetic code editing data from \mymodel{} outperform models trained on original commit or regenerated data by 4.30--10.64\%. Although commit data is structurally aligned, its standardized format and limited complexity reduce effectiveness. In contrast, \mymodel{} produces diverse, realistic (pre-edit, instruction, post-edit) triplets that closely reflect practical usage, resulting in substantial performance gains.}

\subsection{RQ3: Multi-LLM Data Integration}
To investigate whether combining data generated by different LLMs can enhance the performance of a fine-tuned model, we fine-tune Qwen3-8B-Base on datasets consisting of data generated individually by each LLM as well as their combined data.

\subsubsection{Design}
We fine-tune Qwen3-8B-Base with \mydataset{}-DS, \mydataset{}-Qwen3, and \mydataset{}. As described in Section~\ref{sec:data_synthesis}, \mydataset{}-DS and \mydataset{}-Qwen3 are constructed from the data generated by DeepSeek-V3 and Qwen3-Instruct, respectively, representing datasets synthesized from a single model. \mydataset{} is the dataset that integrates data from both models.

\subsubsection{Results}
As shown in Table~\ref{tab:mix_data_two_models}, the models fine-tuned on \mydataset{} consistently achieve superior performance compared with those trained on \mydataset{}-DS or \mydataset{}-Qwen3, providing strong evidence that integrating data synthesized from different LLMs is an effective strategy for enhancing model performance. 

\begin{figure}[htbp]
  \centering
  \begin{subfigure}[t]{0.47\linewidth}
    \centering
    \includegraphics[trim=2cm 1cm 2cm 1cm, clip, width=\linewidth]{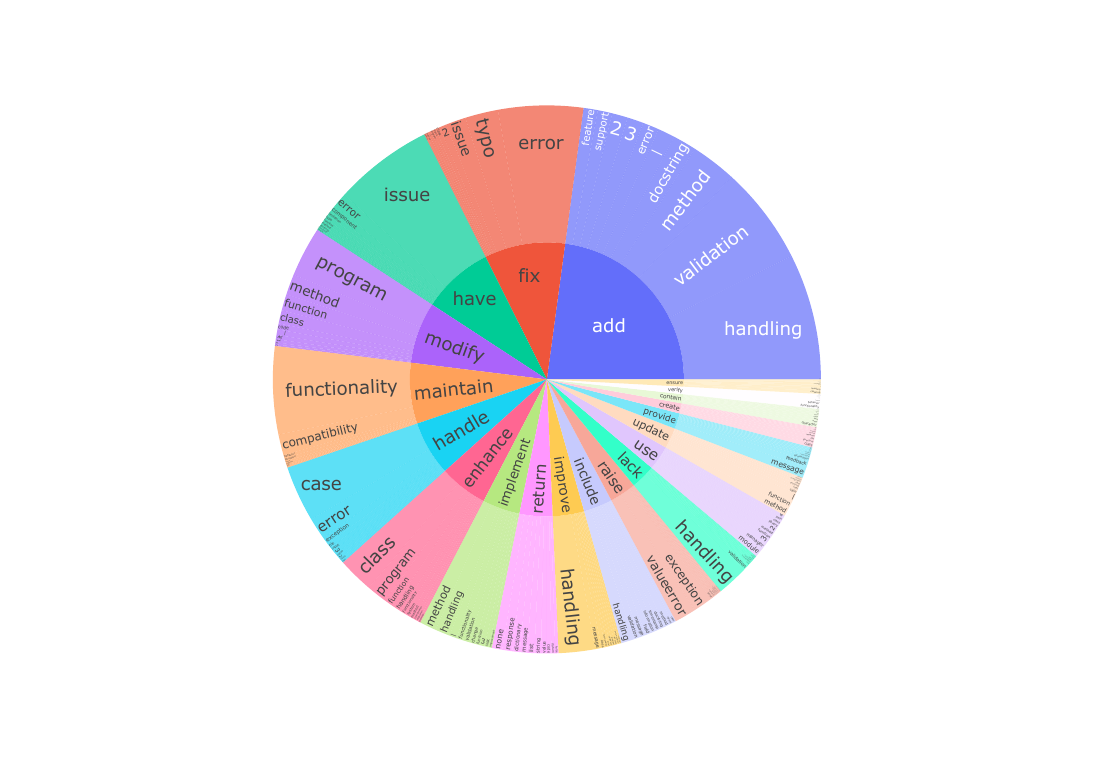}
    \caption{\mydataset{}-DS}
  \end{subfigure}
  \hfill
  \begin{subfigure}[t]{0.47\linewidth}
    \centering
    \includegraphics[trim=2cm 1cm 2cm 1cm, clip, width=\linewidth]{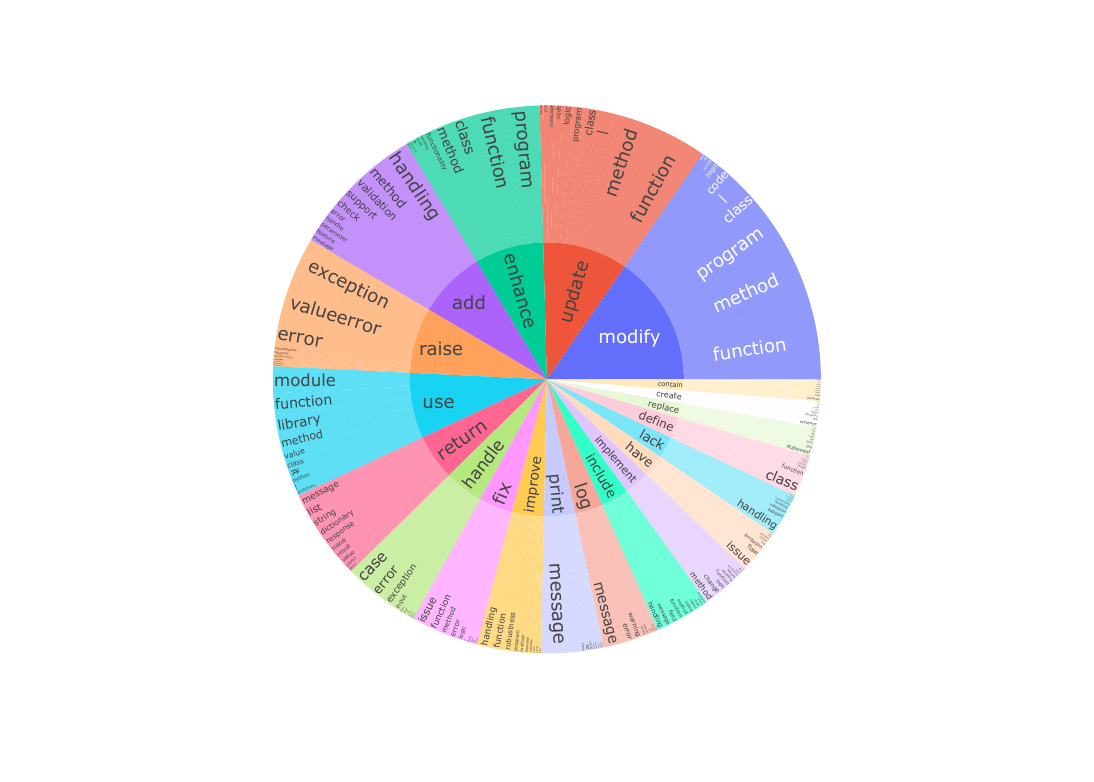}
    \caption{\mydataset{}-Qwen3}
  \end{subfigure}
  \caption{Sunburst plot of the top 20 most frequent verbs, with their corresponding top 10 root nouns, in \mydataset{}-DS and \mydataset{}-Qwen3 datasets.}
  \label{fig:instr_sun_burst}
\end{figure}

To ascertain how editing tasks are distributed in the dataset generated by each model, we illustrate the sunburst plots depicting the most common verbs and their most frequent objects in the \mydataset{}-DS and \mydataset{}-Qwen3 datasets in Figure~\ref{fig:instr_sun_burst}. These verbs and objects reflect the types of editing tasks represented in the edit instructions. As shown in the figure, the two LLMs produce edit instructions with distinct distributions of verb-object phrases. Therefore, combining the edit instructions generated by both LLMs leads to increased diversity in both the types of editing tasks (task diversity) and how these tasks are described (linguistic diversity). This includes variations in phrasing, verb choices, and the level of detail used to specify each task, which contributes to a richer and more flexible set of instructions.

\begin{figure}[htbp]
  \centering
  \begin{subfigure}[t]{0.47\linewidth}
    \centering
    \includegraphics[width=\linewidth]{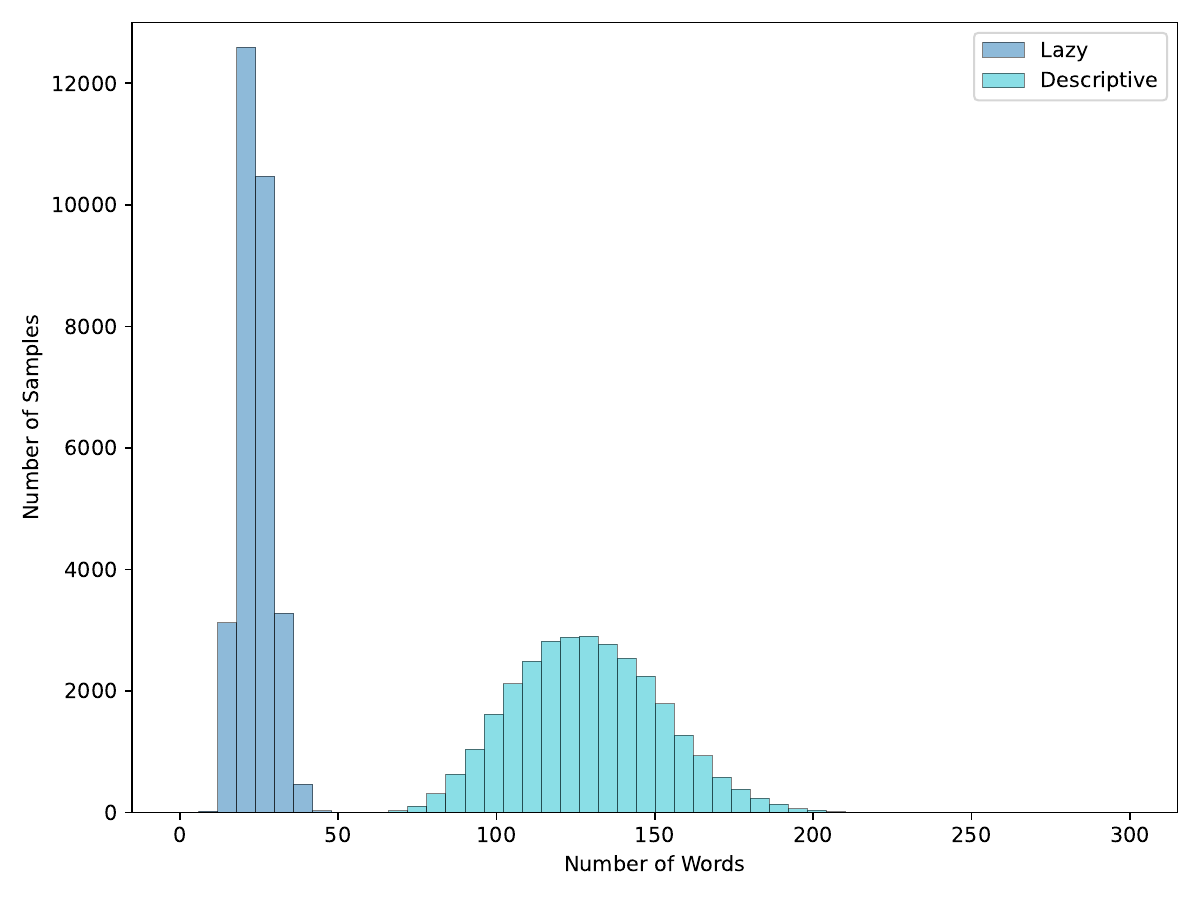}
    \caption{\mydatasetQwen{}-DS}
  \end{subfigure}
  \hfill
  \begin{subfigure}[t]{0.47\linewidth}
    \centering
    \includegraphics[width=\linewidth]{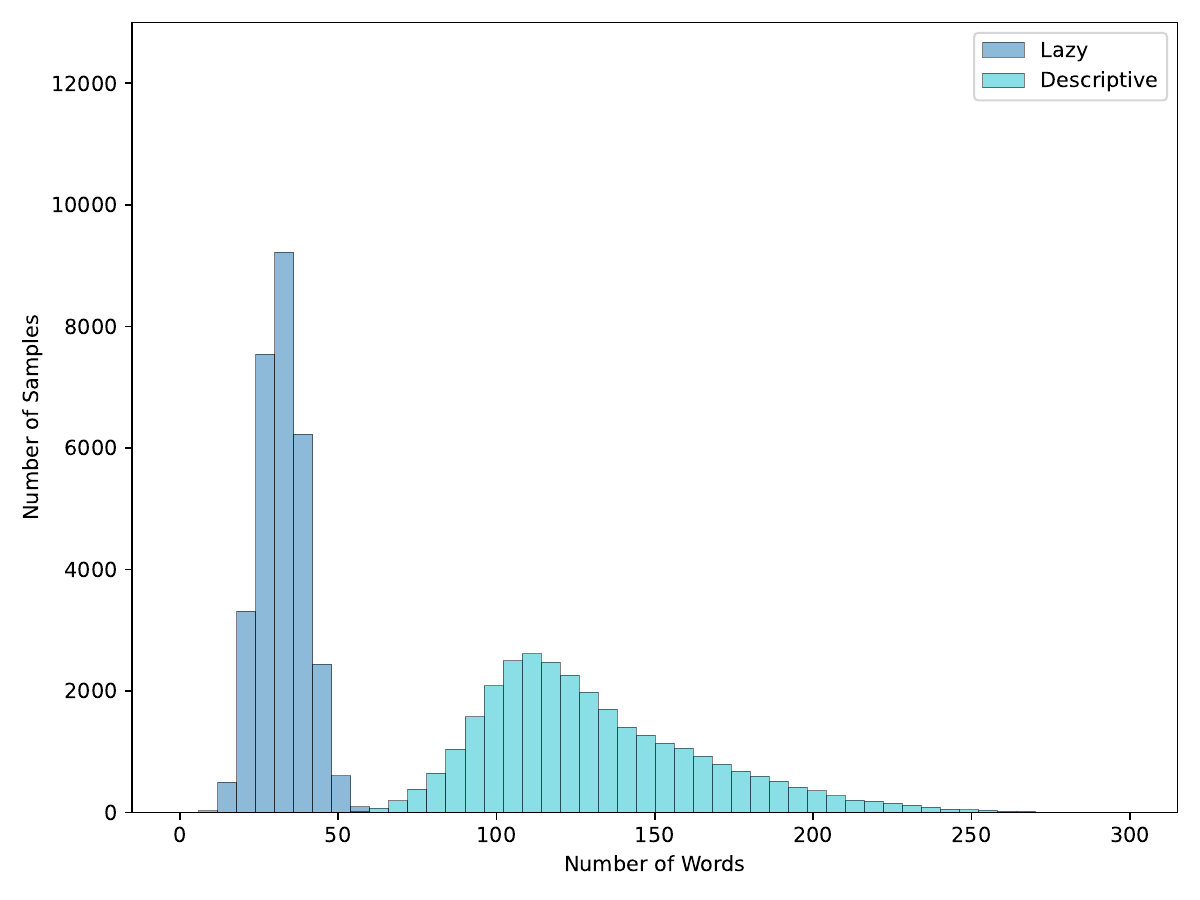}
    \caption{\mydatasetDS{}{}}
  \end{subfigure}
  \caption{Histograms of instruction lengths (in word counts) for descriptive and lazy edit instructions in \mydatasetDS{}{} and \mydatasetQwen{}.}
  \label{fig:text_length}
\end{figure}

Figure~\ref{fig:text_length} presents four histograms that illustrate the distribution of edit instruction lengths, measured in word counts, for both descriptive and lazy edit instructions in \mydataset{}-DS and \mydataset{}-Qwen3. The results reveal distinct patterns across the two datasets: while the lengths of edit instructions in \mydataset{}-DS are more concentrated within a narrower range, those in \mydataset{}-Qwen3 exhibit a broader distribution regardless of the instruction form. These differences stem from the distinct knowledge and generation characteristics of the underlying synthesis models. By combining the two datasets, the distributions are effectively merged, producing a more balanced and diverse set of instructions, which in turn leads to improved downstream results.

\begin{table}[htbp]
    \caption{Pass@1 (\%) of Qwen3-8B-Base fine-tuned on datasets from individual and combined LLM sources.}
    \label{tab:mix_data_two_models}
    \centering
    \scriptsize
    \begin{tabular}{llcccc}
    \toprule
     Base Model &Dataset&Data Amount
& Lazy& Descriptive &Overall
\\
    \midrule
    \multirow{3}{*}{Qwen3-8B-Base} & \mydatasetQwen{}& \multirow{3}{*}{60,000} & 42.29 &54.67 &48.48 
\\
 & \mydatasetDS{}& & 43.10 &55.00 &49.05 
\\
 & \mydataset{}& & \textbf{45.00} & \textbf{58.52} &\textbf{51.76} \\
    \bottomrule
    \end{tabular}
\end{table}

\finding{The mixed dataset, built from multiple LLMs, improves task and language diversity as well as instruction length balance, leading to better fine-tuning performance.}

\subsection{RQ4: Integration of Descriptive and Lazy Instruction Styles}
\label{sec:desc_and_lazy}
To investigate whether integrating both descriptive and lazy styles of edit instructions can promote data diversity and enhance the performance of instruction-tuned models, we fine-tune Qwen3-8B-Base using 20,000 filtered samples from each style in \mydataset{} via \myfiltering{}, along with a mixed dataset \mydatasetft{} formed by combining 10,000 samples from each style. All data volumes are standardized to 20,000 samples to ensure fairness. 

\subsubsection{Design}
We construct three fine-tuning datasets based on \mydataset{} to enable a comprehensive comparison:
\begin{itemize}
    \item \mydescft{}: comprising 20,000 filtered descriptive-style instructions processed using \myfiltering{}.
    \item \mylazyft{}: containing 20,000 filtered lazy-style instructions processed using \myfiltering{}.
    \item \mydatasetft{}: a hybrid dataset created by independently applying \myfiltering{} to the descriptive and lazy portions of \mydataset{}, then combining 10,000 samples from each filtered subset to form a balanced set of 20,000 instances.
\end{itemize}

All datasets are standardized to 20,000 samples to ensure equitable experimental conditions. The Qwen3-8B-Base model is fine-tuned separately on each dataset.

\subsubsection{Results}
The results in Table \ref{tab:desc_and_lazy} demonstrate that fine-tuning on \mydatasetft{} yields the best overall performance, achieving 54.10\% pass@1, with superior results on both lazy and descriptive instructions. This suggests that exposure to diverse instructional formats enhances the model's ability to generalize across different query types.

Interestingly, while each specialized model excels in its respective domain, their performance exhibits notable asymmetry. The model trained exclusively on lazy-style instructions (\mylazyft{}) demonstrates competitive performance on descriptive-style instructions, actually surpassing the performance of the descriptively-trained model on its native instruction style. We hypothesize that this cross-style generalization advantage stems from the inherently challenging nature of lazy-style instructions, which require the model to develop stronger inference and contextual understanding capabilities.

\begin{table}[htbp]
    \caption{Pass@1 (\%) of Qwen3-8B-Base fine-tuned on descriptive, lazy, and mixed instruction styles.}
    \label{tab:desc_and_lazy}
    \centering
    \scriptsize
    \begin{tabular}{lllcccc}
    \toprule
     Base Model & Dataset &Description Type  &Data Amount& Lazy& Descriptive &Overall
\\
    \midrule
    \multirow{3}{*}{Qwen3-8B-Base} & \mydescft{} & Descriptive &\multirow{3}{*}{20,000}& 44.43 & 54.81 
 &49.62 
\\
     & \mylazyft{} & Lazy && 45.00 & 57.19 
 &51.10 
\\
     & \mydatasetft{} & Descriptive + Lazy  && \textbf{48.14} & \textbf{60.05} 
 &\textbf{54.10} \\
    \bottomrule
    \end{tabular}
\end{table}

\finding{Fine-tuning with a mix of descriptive and lazy instructions yields the best performance. Exposure to diverse formats improves generalization, and training on lazy instructions alone also supports effective cross-style performance due to their demanding nature.}

\subsection{RQ5: Impact of \myfiltering{} on Data Quality Enhancement}
\label{sec:filter_eval}
To evaluate the contribution of our proposed \myfiltering{} method to the \mymodel{} pipeline, we fine-tune Qwen3-8B-Base using both filtered and randomly sampled datasets---each containing 20,000 balanced instruction-style samples---and compare their performance via pass@1 scores.

\subsubsection{Design}
We apply the \myfiltering{} method to separately filter descriptive-style and lazy-style instructions from three original datasets: \mydatasetQwen{}, \mydatasetDS{}, and \mydataset{}. For each dataset, we extract exactly 10,000 descriptive-style and 10,000 lazy-style instructions after filtering, and then combine them to form three filtered datasets: \mydatasetQwenft{}, \mydatasetDSft{}, and \mydatasetft{}, each containing exactly 20,000 samples with balanced style distribution.

To provide comprehensive baseline comparisons, we construct two types of contrastive datasets. First, we create an alternative filtering baseline by separately applying \myfiltering{} to descriptive-style and lazy-style instructions in \mydatasetQwen{} and \mydatasetDS{}, extracting exactly 5,000 descriptive-style and 5,000 lazy-style instructions from each dataset, and then merging these to form OCEPack-QD-FT (totaling 20,000 samples with 10,000 descriptive-style and 10,000 lazy-style instructions). Second, we separately randomly select 10,000 descriptive-style and 10,000 lazy-style instructions from each of the three original datasets and combine them to construct randomly-sampled datasets: OCEPack-Qwen3-Rand, OCEPack-DS-Rand, and OCEPack-Rand, each containing 20,000 samples with balanced style distribution.

All seven datasets are used to independently instruction-tune the base model Qwen3-8B-Base. The performance of each resulting fine-tuned model is evaluated and compared based on the pass@1 score.

Furthermore, to investigate whether data filtering could enable more efficient training—achieving better results with fewer data—we fine-tuned the Qwen3-8B-Base model on subsets of \mydataset{}, \mydatasetft{}, and \commitpackft{}.

\subsubsection{Results}
The experimental results in Table~\ref{tab:filtering} highlight the effectiveness of the \myfiltering{} method, revealing substantial performance differences across filtering strategies.

Across all settings, filtered datasets consistently outperformed their randomly sampled counterparts. Our main contribution, \mydatasetft{}, constructed by filtering the combined DS+Qwen3 data, achieved the highest overall pass@1 score of 54.10\%. By contrast, the alternative strategy of filtering before merging (\mydatasetQDft{}) yielded a slightly lower overall score of 53.57\%. Notably, however, the \mydatasetQDft{} model obtained marginally better results on lazy-style instructions. This trade-off suggests that filtering after dataset integration enhances overall data quality and thematic coverage, even if it sacrifices minor performance on a single instruction style. Considering both effectiveness and implementation simplicity, we adopt the mix-then-filter approach.

The benefits of the method are not limited to the combined dataset. When applied separately to Qwen3- and DeepSeek-generated data, filtering raised the overall scores by 1.67\% and 1.42\%, respectively. These consistent gains across distinct LLM sources underscore the robustness and generalizability of the \myfiltering{} approach.

As shown in Table~\ref{tab:quantity_vs_quality}, the filtered dataset \mydatasetft{} (20k samples) outperforms the much larger unfiltered dataset \mydataset{} (60k samples), yielding a 2.34\% absolute gain in overall pass@1. This result indicates that \myfiltering{} reduces the amount of training data required to reach superior performance by removing redundant data and filtering out noisy instances. We also compare with the model fine-tuned with the whole Python subset of \commitpackft{}. Despite containing a large volume of data (56k samples), \commitpackft{} yielded only 20.98\% in overall pass@1, far below both \mydataset{} and \mydatasetft{}. This observation is consistent with our earlier finding that commit data fails to effectively enhance LLM performance on code editing tasks, even when the amount of data is increased. 

\begin{table}[htbp]
    \caption{Comparison of pass@1 (\%) scores across instruction-tuned models using datasets processed with different filtering strategies.}
    \label{tab:filtering}
    \centering
    \scriptsize
    \begin{tabular}{lllcccc}
    \toprule
     Base Model  &Dataset& Composition of Data &Data Amount& Lazy& Descriptive &Overall
\\
    \midrule
     \multirow{7}{*}{Qwen3-8B-Base}&\mydatasetQwenrand{}&  Qwen3 & \multirow{2}{*}{20,000} & 44.38 & 54.43 
 &49.40 
\\
      &\mydatasetQwenft{}& Qwen3$^\dagger$  & & 45.33 & 56.81 
 &51.07 
\\
\cmidrule(lr){2-7}
    &\mydatasetDSrand{}&  DS & \multirow{2}{*}{20,000} & 40.81 & 57.14 
 &48.98 
\\
    &\mydatasetDSft{}& DS$^\dagger$ & & 43.67 &57.14 
 &50.40 
\\
\cmidrule(lr){2-7}
    &\mydatasetrand{}& DS + Qwen3 & \multirow{3}{*}{20,000} & 46.71 &54.19 
 &50.45 
\\
    & \mydatasetQDft{} & DS$^\dagger$ + Qwen3$^\dagger$ & & 49.52 &57.62 
 &53.57 
\\
    &\mydatasetft{}& (DS + Qwen3)$^\ddagger$ & & 48.14 &60.05 
 &54.10 \\
    \bottomrule
    \multicolumn{7}{l}{$^{\dagger}$Data processed with the \myfiltering{} method.} \\
    \multicolumn{7}{l}{$^{\ddagger}$DS and Qwen3 data were first combined and then filtered with the \myfiltering{} method.} 
    \end{tabular}
\end{table}

\begin{table}[htbp]
    \caption{Pass@1 (\%) of fine-tuned models on various datasets.}
    \label{tab:quantity_vs_quality}
    \centering
    \scriptsize
    \begin{tabular}{llcccc}
    \toprule
     Base Model & Dataset Name&Data Amount
& Lazy& Descriptive &Overall
\\
    \midrule
     \multirow{3}{*}{Qwen3-8B-Base} & \mydataset{}& 60000
& 45.00 & 58.52 
 &51.76 
\\
     & \commitpackft{}& 56009
& 18.14 & 23.81 
 &20.98 
\\
     & \mydatasetft{}& 20000
& 48.14 & 60.05 
 &54.10 \\
    \bottomrule
    \end{tabular}
\end{table}

\finding{\myfiltering{} demonstrates a “less-is-more” effect: despite reducing the dataset size by two-thirds, it yields superior fine-tuning performance. This is achieved by systematically removing redundant and noisy samples, thereby improving data quality, reducing training overhead, and boosting fine-tuning efficiency.}

\section{Related Work}
\subsection{Code Editing}
The task of automated code editing was a vibrant research area long before the advent of modern Large Language Models. Pre-LLM approaches focused on learning from historical data, either by applying mined fix templates to repair bugs~\cite{kim2013automatic} or by treating edits as a translation task using neural models on token sequences~\cite{tufano2019learning} and Abstract Syntax Trees (ASTs)~\cite{chakraborty2020codit}. Other research tackled more specific scenarios, such as automating API adaptation and recommendation~\cite{nguyen2010graph, nguyen2016api} or interactively completing refactorings within the IDE~\cite{foster2012witchdoctor}.

The advent of Large Language Models introduced a powerful new paradigm, though initial research was predominantly focused on code generation. Foundational models like OpenAI's Codex established benchmarks for synthesizing code from natural language prompts~\cite{chen2021evaluating}, while others like AlphaCode and Code Llama pushed the boundaries of algorithmic problem-solving and open-source capabilities~\cite{li2022competition, roziere2023code}. Collectively, this line of work established LLMs as powerful tools for creating new code from scratch~\cite{zhang2023survey}. However, empirical studies consistently show that software engineers spend the majority of their time modifying existing code, highlighting a gap between the research focus and practical developer needs~\cite{nguyen2013study, mozannar2024reading}.

This reality has motivated a growing body of research on applying LLMs specifically to code editing. Previous studies have explored this task from several angles. A significant portion of work has focused on bug fixing, a critical subset of code editing, with numerous studies evaluating LLMs on this task~\cite{zhang2023self, moon2023coffee, shinn2023reflexion, chen2023teaching, olausson2023self, jin2023inferfix, joshi2023repair, wei2023copiloting, li2022automating}. Another line of research has focused on the fill-in-the-middle inference strategy, where models like InCoder or Code Llama complete code at specific insertion points, rather than following a natural language directive~\cite{bavarian2022efficient, fried2022incoder, roziere2023code,guo2024deepseek}. More recently, the field has progressed towards true instructional code editing, where models are guided by natural language commands. This training paradigm is exemplified by datasets like Octopack, which frames Git commits as natural examples of code edits paired with human instructions~\cite{muennighoff2023octopack}. Such datasets have fueled the development of specialized models like Coeditor, which leverages repo-level context for multi-round editing~\cite{wei2024coeditor}, EDITLORD, which focuses on learning a diverse set of edit operations from user instructions~\cite{li2025editlord}, InstructCoder, explicitly trained for instructional tasks~\cite{hu2023instructcoder}, and dedicated benchmarks like CanItEdit, which provides a suite of hand-crafted instructional editing problems~\cite{cassano2024can}. These methods all leverage large corpora of Git commits, using commit messages or PR descriptions as proxies for natural language instructions and code diffs as the ground-truth changes.

While these methods demonstrate the feasibility of learning from historical commits, they are fundamentally limited by the inherent quality and style of that data. Our work, \mymodel{}, proposes a distinct paradigm that shifts from curation to creation. Instead of using historical commits as direct training examples, we leverage open-source code snippets as seeds to synthesize entirely new, high-quality instruction data tailored specifically for complex code editing tasks.

\subsection{Data Synthesis for Large Language Models}
The performance of modern LLMs is deeply tied to the quality of the data used for Supervised Fine-Tuning (SFT), or instruction tuning. As creating high-quality, large-scale instruction datasets manually is a laborious and expensive process, generating synthetic data via LLMs themselves has emerged as a key enabling technique. This approach, often referred to as post-training alignment, has become a standard for enhancing model capabilities~\cite{wei2024selfcodealign, wang-etal-2023-self-instruct, xu2024wizardlm, luo2024wizardcoder, zhao2024self, li2024synthetic}.

These methods have been extended to the code domain, primarily to enhance code generation models. For example, WizardCoder employs the Evol-Instruct framework to enhance the complexity of instructions in the CodeAlpaca dataset, which was itself synthetically generated using the Alpaca and Self-Instruct pipelines~\cite{luo2024wizardcoder, xu2024wizardlm, chaudhary2023code, taori2023stanford, wang-etal-2023-self-instruct}. A more recent class of methods conditions the generation process on seed code derived from real code files. Approaches such as Self-CodeAlign~\cite{wei2024selfcodealign}, WaveCoder~\cite{yu2023wavecoder}, and our own inspiration, OSS-Instruct~\cite{wei2024magicoder}, belong to this category, significantly improving task diversity by leveraging real-world code.

Explicit work on synthesizing data for code editing is much rarer. InstructCoder, for instance, is one of the few prior methods explicitly designed to generate synthetic data for code editing tasks by conditioning on seed examples~\cite{hu2023instructcoder}. Nevertheless, even these advanced methods have focused primarily on function-level editing snippets, and have not been widely tested on generating file-level examples that include multiple classes or functions.

While prior work has focused on data synthesis for code generation, instruction-guided code editing remains underexplored. \mymodel{} addresses this by generating novel (pre-edit, instruction, post-edit) triplets that better reflect real-world editing needs.

\section{Threats to Validity}
We identify potential threats to the validity of our study and describe mitigation strategies.

\parabf{Internal Validity}
Results may be influenced by LLM selection, hyperparameters, or random seeds. We mitigate this by evaluating multiple open-source models with consistent settings and repeated runs. Bias in synthetic data is another concern, which we address by using two complementary LLMs, generating both \emph{lazy} and \emph{descriptive} instructions, and applying diff- and topic-based filtering to enhance diversity and reduce redundancy.

\parabf{Construct Validity}
Evaluation metrics and benchmarks may not fully capture real-world code editing scenarios. We use established code editing benchmarks~\cite{cassano2024can} and include tasks of varying complexity in our synthetic dataset to better reflect practical editing challenges.

\parabf{External Validity}
Findings may be limited to the specific models, languages, and code domains studied. \mymodel{} is adaptable to other domains by replacing seed code snippets, but performance gains may vary for proprietary models or highly specialized codebases.

\parabf{Conclusion Validity}
Performance improvements may depend on dataset size, instruction style mix, and filtering thresholds. We mitigate this by conducting ablation studies (RQ3–RQ5) to verify that gains are robust across configurations.

\section{Conclusions}
This paper presented~\mymodel{}, an open-source pipeline for synthesizing instruction-tuning data for code editing. By combining multiple open-source LLMs and applying \myfiltering{}, the pipeline produces realistic edit triplets and ensures both quality and diversity. Based on this pipeline, we constructed \mydatasetft{}, a lightweight dataset of 20,000 samples. Experiments on the \canitedit{} benchmark demonstrate that models fine-tuned on \mydatasetft{} achieve consistent improvements in pass@1, with relative gains between 4.50\% and 20.79\% over existing instruction-tuned counterparts. These results confirm the effectiveness of task-aligned synthetic data in enhancing LLMs’ code editing capabilities without relying on proprietary resources. Future work will focus on broadening the coverage of programming languages and addressing more complex editing scenarios, further advancing open-source research in instruction-guided code editing.

\section{Data Availability}
\label{sec:open-science}
To facilitate the replication study, we have released our data and code at~\url{https://github.com/zkzhang88/OpenCodeEdit-public-1}. 

\bibliographystyle{ACM-Reference-Format}

\begin{thebibliography}{82}


\ifx \showCODEN    \undefined \def \showCODEN     #1{\unskip}     \fi
\ifx \showISBNx    \undefined \def \showISBNx     #1{\unskip}     \fi
\ifx \showISBNxiii \undefined \def \showISBNxiii  #1{\unskip}     \fi
\ifx \showISSN     \undefined \def \showISSN      #1{\unskip}     \fi
\ifx \showLCCN     \undefined \def \showLCCN      #1{\unskip}     \fi
\ifx \shownote     \undefined \def \shownote      #1{#1}          \fi
\ifx \showarticletitle \undefined \def \showarticletitle #1{#1}   \fi
\ifx \showURL      \undefined \def \showURL       {\relax}        \fi
\providecommand\bibfield[2]{#2}
\providecommand\bibinfo[2]{#2}
\providecommand\natexlab[1]{#1}
\providecommand\showeprint[2][]{arXiv:#2}

\bibitem[{Alibaba Cloud}(2024)]%
        {alibaba_cloud_2024}
\bibfield{author}{\bibinfo{person}{{Alibaba Cloud}}.} \bibinfo{year}{2024}\natexlab{}.
\newblock \bibinfo{title}{Model Studio - Alibaba Cloud}.
\newblock \bibinfo{howpublished}{\url{https://help.aliyun.com/zh/model-studio/models}}.
\newblock


\bibitem[Almeida et~al\mbox{.}(2024)]%
        {almeida2024automatic}
\bibfield{author}{\bibinfo{person}{Aylton Almeida}, \bibinfo{person}{Laerte Xavier}, {and} \bibinfo{person}{Marco~Tulio Valente}.} \bibinfo{year}{2024}\natexlab{}.
\newblock \showarticletitle{Automatic library migration using large language models: First results}. In \bibinfo{booktitle}{\emph{Proceedings of the 18th ACM/IEEE International Symposium on Empirical Software Engineering and Measurement}}. \bibinfo{pages}{427--433}.
\newblock


\bibitem[Austin et~al\mbox{.}(2021)]%
        {austin2021program}
\bibfield{author}{\bibinfo{person}{Jacob Austin}, \bibinfo{person}{Augustus Odena}, \bibinfo{person}{Maxwell Nye}, \bibinfo{person}{Maarten Bosma}, \bibinfo{person}{Henryk Michalewski}, \bibinfo{person}{David Dohan}, \bibinfo{person}{Ellen Jiang}, \bibinfo{person}{Carrie Cai}, \bibinfo{person}{Michael Terry}, \bibinfo{person}{Quoc Le}, {et~al\mbox{.}}} \bibinfo{year}{2021}\natexlab{}.
\newblock \showarticletitle{Program synthesis with large language models}.
\newblock \bibinfo{journal}{\emph{arXiv preprint arXiv:2108.07732}} (\bibinfo{year}{2021}).
\newblock


\bibitem[Bavarian et~al\mbox{.}(2022)]%
        {bavarian2022efficient}
\bibfield{author}{\bibinfo{person}{Mohammad Bavarian}, \bibinfo{person}{Heewoo Jun}, \bibinfo{person}{Nikolas Tezak}, \bibinfo{person}{John Schulman}, \bibinfo{person}{Christine McLeavey}, \bibinfo{person}{Jerry Tworek}, {and} \bibinfo{person}{Mark Chen}.} \bibinfo{year}{2022}\natexlab{}.
\newblock \showarticletitle{Efficient training of language models to fill in the middle}.
\newblock \bibinfo{journal}{\emph{arXiv preprint arXiv:2207.14255}} (\bibinfo{year}{2022}).
\newblock


\bibitem[Carissimi et~al\mbox{.}(2025)]%
        {carissimi2025towards}
\bibfield{author}{\bibinfo{person}{Michele Carissimi}, \bibinfo{person}{Martina Saletta}, {and} \bibinfo{person}{Claudio Ferretti}.} \bibinfo{year}{2025}\natexlab{}.
\newblock \showarticletitle{Towards Leveraging Large Language Model Summaries for Topic Modeling in Source Code}.
\newblock \bibinfo{journal}{\emph{arXiv preprint arXiv:2504.17426}} (\bibinfo{year}{2025}).
\newblock


\bibitem[Cassano et~al\mbox{.}(2024)]%
        {cassano2024can}
\bibfield{author}{\bibinfo{person}{Federico Cassano}, \bibinfo{person}{Luisa Li}, \bibinfo{person}{Akul Sethi}, \bibinfo{person}{Noah Shinn}, \bibinfo{person}{Abby Brennan-Jones}, \bibinfo{person}{Jacob Ginesin}, \bibinfo{person}{Edward Berman}, \bibinfo{person}{George Chakhnashvili}, \bibinfo{person}{Anton Lozhkov}, \bibinfo{person}{Carolyn~Jane Anderson}, {et~al\mbox{.}}} \bibinfo{year}{2024}\natexlab{}.
\newblock \showarticletitle{Can It Edit? Evaluating the Ability of Large Language Models to Follow Code Editing Instructions}. In \bibinfo{booktitle}{\emph{First Conference on Language Modeling}}.
\newblock


\bibitem[Chakraborty et~al\mbox{.}(2020)]%
        {chakraborty2020codit}
\bibfield{author}{\bibinfo{person}{Saikat Chakraborty}, \bibinfo{person}{Yangruibo Ding}, \bibinfo{person}{Miltiadis Allamanis}, {and} \bibinfo{person}{Baishakhi Ray}.} \bibinfo{year}{2020}\natexlab{}.
\newblock \showarticletitle{Codit: Code editing with tree-based neural models}.
\newblock \bibinfo{journal}{\emph{IEEE Transactions on Software Engineering}} \bibinfo{volume}{48}, \bibinfo{number}{4} (\bibinfo{year}{2020}), \bibinfo{pages}{1385--1399}.
\newblock


\bibitem[Chaudhary(2023)]%
        {chaudhary2023code}
\bibfield{author}{\bibinfo{person}{Sahil Chaudhary}.} \bibinfo{year}{2023}\natexlab{}.
\newblock \bibinfo{title}{Code alpaca: An instruction-following llama model for code generation}.
\newblock


\bibitem[Chen et~al\mbox{.}(2024)]%
        {chen2024alpagasus}
\bibfield{author}{\bibinfo{person}{Lichang Chen}, \bibinfo{person}{Shiyang Li}, \bibinfo{person}{Jun Yan}, \bibinfo{person}{Hai Wang}, \bibinfo{person}{Kalpa Gunaratna}, \bibinfo{person}{Vikas Yadav}, \bibinfo{person}{Zheng Tang}, \bibinfo{person}{Vijay Srinivasan}, \bibinfo{person}{Tianyi Zhou}, \bibinfo{person}{Heng Huang}, {and} \bibinfo{person}{Hongxia Jin}.} \bibinfo{year}{2024}\natexlab{}.
\newblock \showarticletitle{AlpaGasus: Training a Better Alpaca with Fewer Data}. In \bibinfo{booktitle}{\emph{The Twelfth International Conference on Learning Representations}}.
\newblock
\urldef\tempurl%
\url{https://openreview.net/forum?id=FdVXgSJhvz}
\showURL{%
\tempurl}


\bibitem[Chen et~al\mbox{.}(2021)]%
        {chen2021evaluating}
\bibfield{author}{\bibinfo{person}{Mark Chen}, \bibinfo{person}{Jerry Tworek}, \bibinfo{person}{Heewoo Jun}, \bibinfo{person}{Qiming Yuan}, \bibinfo{person}{Henrique Ponde De~Oliveira Pinto}, \bibinfo{person}{Jared Kaplan}, \bibinfo{person}{Harri Edwards}, \bibinfo{person}{Yuri Burda}, \bibinfo{person}{Nicholas Joseph}, \bibinfo{person}{Greg Brockman}, {et~al\mbox{.}}} \bibinfo{year}{2021}\natexlab{}.
\newblock \showarticletitle{Evaluating large language models trained on code}.
\newblock \bibinfo{journal}{\emph{arXiv preprint arXiv:2107.03374}} (\bibinfo{year}{2021}).
\newblock


\bibitem[Chen et~al\mbox{.}(2023)]%
        {chen2023teaching}
\bibfield{author}{\bibinfo{person}{Xinyun Chen}, \bibinfo{person}{Maxwell Lin}, \bibinfo{person}{Nathanael Sch{\"a}rli}, {and} \bibinfo{person}{Denny Zhou}.} \bibinfo{year}{2023}\natexlab{}.
\newblock \showarticletitle{Teaching large language models to self-debug}.
\newblock \bibinfo{journal}{\emph{arXiv preprint arXiv:2304.05128}} (\bibinfo{year}{2023}).
\newblock


\bibitem[Cummins et~al\mbox{.}(2024)]%
        {cummins2024don}
\bibfield{author}{\bibinfo{person}{Chris Cummins}, \bibinfo{person}{Volker Seeker}, \bibinfo{person}{Jordi Armengol-Estap{\'e}}, \bibinfo{person}{Aram~H Markosyan}, \bibinfo{person}{Gabriel Synnaeve}, {and} \bibinfo{person}{Hugh Leather}.} \bibinfo{year}{2024}\natexlab{}.
\newblock \showarticletitle{Don't Transform the Code, Code the Transforms: Towards Precise Code Rewriting using LLMs}.
\newblock \bibinfo{journal}{\emph{arXiv preprint arXiv:2410.08806}} (\bibinfo{year}{2024}).
\newblock


\bibitem[{DeepSeek}(2024)]%
        {deepseek_platform_2024}
\bibfield{author}{\bibinfo{person}{{DeepSeek}}.} \bibinfo{year}{2024}\natexlab{}.
\newblock \bibinfo{title}{DeepSeek Platform}.
\newblock \bibinfo{howpublished}{\url{https://platform.deepseek.com/}}.
\newblock


\bibitem[Deng et~al\mbox{.}(2025)]%
        {deng2025nocode}
\bibfield{author}{\bibinfo{person}{Le Deng}, \bibinfo{person}{Zhonghao Jiang}, \bibinfo{person}{Jialun Cao}, \bibinfo{person}{Michael Pradel}, {and} \bibinfo{person}{Zhongxin Liu}.} \bibinfo{year}{2025}\natexlab{}.
\newblock \showarticletitle{NoCode-bench: A Benchmark for Evaluating Natural Language-Driven Feature Addition}.
\newblock \bibinfo{journal}{\emph{arXiv preprint arXiv:2507.18130}} (\bibinfo{year}{2025}).
\newblock


\bibitem[Foster et~al\mbox{.}(2012)]%
        {foster2012witchdoctor}
\bibfield{author}{\bibinfo{person}{Stephen~R Foster}, \bibinfo{person}{William~G Griswold}, {and} \bibinfo{person}{Sorin Lerner}.} \bibinfo{year}{2012}\natexlab{}.
\newblock \showarticletitle{WitchDoctor: IDE support for real-time auto-completion of refactorings}. In \bibinfo{booktitle}{\emph{2012 34th international conference on software engineering (ICSE)}}. IEEE, \bibinfo{pages}{222--232}.
\newblock


\bibitem[Fried et~al\mbox{.}(2022)]%
        {fried2022incoder}
\bibfield{author}{\bibinfo{person}{Daniel Fried}, \bibinfo{person}{Armen Aghajanyan}, \bibinfo{person}{Jessy Lin}, \bibinfo{person}{Sida Wang}, \bibinfo{person}{Eric Wallace}, \bibinfo{person}{Freda Shi}, \bibinfo{person}{Ruiqi Zhong}, \bibinfo{person}{Wen-tau Yih}, \bibinfo{person}{Luke Zettlemoyer}, {and} \bibinfo{person}{Mike Lewis}.} \bibinfo{year}{2022}\natexlab{}.
\newblock \showarticletitle{Incoder: A generative model for code infilling and synthesis}.
\newblock \bibinfo{journal}{\emph{arXiv preprint arXiv:2204.05999}} (\bibinfo{year}{2022}).
\newblock


\bibitem[Gunasekar et~al\mbox{.}(2023)]%
        {gunasekar2023textbooks}
\bibfield{author}{\bibinfo{person}{Suriya Gunasekar}, \bibinfo{person}{Yi Zhang}, \bibinfo{person}{Jyoti Aneja}, \bibinfo{person}{Caio C{\'e}sar~Teodoro Mendes}, \bibinfo{person}{Allie Del~Giorno}, \bibinfo{person}{Sivakanth Gopi}, \bibinfo{person}{Mojan Javaheripi}, \bibinfo{person}{Piero Kauffmann}, \bibinfo{person}{Gustavo de Rosa}, \bibinfo{person}{Olli Saarikivi}, {et~al\mbox{.}}} \bibinfo{year}{2023}\natexlab{}.
\newblock \showarticletitle{Textbooks are all you need}.
\newblock \bibinfo{journal}{\emph{arXiv preprint arXiv:2306.11644}} (\bibinfo{year}{2023}).
\newblock


\bibitem[Guo et~al\mbox{.}(2024b)]%
        {guo2024deepseek}
\bibfield{author}{\bibinfo{person}{Daya Guo}, \bibinfo{person}{Qihao Zhu}, \bibinfo{person}{Dejian Yang}, \bibinfo{person}{Zhenda Xie}, \bibinfo{person}{Kai Dong}, \bibinfo{person}{Wentao Zhang}, \bibinfo{person}{Guanting Chen}, \bibinfo{person}{Xiao Bi}, \bibinfo{person}{Yu Wu}, \bibinfo{person}{YK Li}, {et~al\mbox{.}}} \bibinfo{year}{2024}\natexlab{b}.
\newblock \showarticletitle{DeepSeek-Coder: When the Large Language Model Meets Programming--The Rise of Code Intelligence}.
\newblock \bibinfo{journal}{\emph{arXiv preprint arXiv:2401.14196}} (\bibinfo{year}{2024}).
\newblock


\bibitem[Guo et~al\mbox{.}(2024a)]%
        {guo2024codeeditorbench}
\bibfield{author}{\bibinfo{person}{Jiawei Guo}, \bibinfo{person}{Ziming Li}, \bibinfo{person}{Xueling Liu}, \bibinfo{person}{Kaijing Ma}, \bibinfo{person}{Tianyu Zheng}, \bibinfo{person}{Zhouliang Yu}, \bibinfo{person}{Ding Pan}, \bibinfo{person}{Yizhi Li}, \bibinfo{person}{Ruibo Liu}, \bibinfo{person}{Yue Wang}, {et~al\mbox{.}}} \bibinfo{year}{2024}\natexlab{a}.
\newblock \showarticletitle{Codeeditorbench: Evaluating code editing capability of large language models}.
\newblock \bibinfo{journal}{\emph{arXiv preprint arXiv:2404.03543}} (\bibinfo{year}{2024}).
\newblock


\bibitem[Guo et~al\mbox{.}(2025)]%
        {guo2025omnigirl}
\bibfield{author}{\bibinfo{person}{Lianghong Guo}, \bibinfo{person}{Wei Tao}, \bibinfo{person}{Runhan Jiang}, \bibinfo{person}{Yanlin Wang}, \bibinfo{person}{Jiachi Chen}, \bibinfo{person}{Xilin Liu}, \bibinfo{person}{Yuchi Ma}, \bibinfo{person}{Mingzhi Mao}, \bibinfo{person}{Hongyu Zhang}, {and} \bibinfo{person}{Zibin Zheng}.} \bibinfo{year}{2025}\natexlab{}.
\newblock \showarticletitle{Omnigirl: A multilingual and multimodal benchmark for github issue resolution}.
\newblock \bibinfo{journal}{\emph{Proceedings of the ACM on Software Engineering}} \bibinfo{volume}{2}, \bibinfo{number}{ISSTA} (\bibinfo{year}{2025}), \bibinfo{pages}{24--46}.
\newblock


\bibitem[Gupta et~al\mbox{.}(2023)]%
        {gupta2023grace}
\bibfield{author}{\bibinfo{person}{Priyanshu Gupta}, \bibinfo{person}{Avishree Khare}, \bibinfo{person}{Yasharth Bajpai}, \bibinfo{person}{Saikat Chakraborty}, \bibinfo{person}{Sumit Gulwani}, \bibinfo{person}{Aditya Kanade}, \bibinfo{person}{Arjun Radhakrishna}, \bibinfo{person}{Gustavo Soares}, {and} \bibinfo{person}{Ashish Tiwari}.} \bibinfo{year}{2023}\natexlab{}.
\newblock \showarticletitle{Grace: Generation using associated code edits}.
\newblock \bibinfo{journal}{\emph{arXiv preprint arXiv:2305.14129}} (\bibinfo{year}{2023}).
\newblock


\bibitem[Hu et~al\mbox{.}(2023)]%
        {hu2023instructcoder}
\bibfield{author}{\bibinfo{person}{Qisheng Hu}, \bibinfo{person}{Kaixin Li}, \bibinfo{person}{Xu Zhao}, \bibinfo{person}{Yuxi Xie}, \bibinfo{person}{Tiedong Liu}, \bibinfo{person}{Hui Chen}, \bibinfo{person}{Qizhe Xie}, {and} \bibinfo{person}{Junxian He}.} \bibinfo{year}{2023}\natexlab{}.
\newblock \showarticletitle{Instructcoder: Empowering language models for code editing}.
\newblock \bibinfo{journal}{\emph{CoRR}} (\bibinfo{year}{2023}).
\newblock


\bibitem[Hu et~al\mbox{.}(2021)]%
        {hu2021model}
\bibfield{author}{\bibinfo{person}{Xia Hu}, \bibinfo{person}{Lingyang Chu}, \bibinfo{person}{Jian Pei}, \bibinfo{person}{Weiqing Liu}, {and} \bibinfo{person}{Jiang Bian}.} \bibinfo{year}{2021}\natexlab{}.
\newblock \showarticletitle{Model complexity of deep learning: A survey}.
\newblock \bibinfo{journal}{\emph{Knowledge and Information Systems}} \bibinfo{volume}{63}, \bibinfo{number}{10} (\bibinfo{year}{2021}), \bibinfo{pages}{2585--2619}.
\newblock


\bibitem[Hui et~al\mbox{.}(2024)]%
        {hui2024qwen2}
\bibfield{author}{\bibinfo{person}{Binyuan Hui}, \bibinfo{person}{Jian Yang}, \bibinfo{person}{Zeyu Cui}, \bibinfo{person}{Jiaxi Yang}, \bibinfo{person}{Dayiheng Liu}, \bibinfo{person}{Lei Zhang}, \bibinfo{person}{Tianyu Liu}, \bibinfo{person}{Jiajun Zhang}, \bibinfo{person}{Bowen Yu}, \bibinfo{person}{Keming Lu}, {et~al\mbox{.}}} \bibinfo{year}{2024}\natexlab{}.
\newblock \showarticletitle{Qwen2. 5-coder technical report}.
\newblock \bibinfo{journal}{\emph{arXiv preprint arXiv:2409.12186}} (\bibinfo{year}{2024}).
\newblock


\bibitem[Iammarino et~al\mbox{.}(2020)]%
        {iammarino2020topic}
\bibfield{author}{\bibinfo{person}{Martina Iammarino}, \bibinfo{person}{Lerina Aversano}, \bibinfo{person}{Mario~Luca Bernardi}, {and} \bibinfo{person}{Marta Cimitile}.} \bibinfo{year}{2020}\natexlab{}.
\newblock \showarticletitle{A topic modeling approach to evaluate the comments consistency to source code}. In \bibinfo{booktitle}{\emph{2020 International Joint Conference on Neural Networks (IJCNN)}}. IEEE, \bibinfo{pages}{1--8}.
\newblock


\bibitem[Jin et~al\mbox{.}(2023)]%
        {jin2023inferfix}
\bibfield{author}{\bibinfo{person}{Matthew Jin}, \bibinfo{person}{Syed Shahriar}, \bibinfo{person}{Michele Tufano}, \bibinfo{person}{Xin Shi}, \bibinfo{person}{Shuai Lu}, \bibinfo{person}{Neel Sundaresan}, {and} \bibinfo{person}{Alexey Svyatkovskiy}.} \bibinfo{year}{2023}\natexlab{}.
\newblock \showarticletitle{Inferfix: End-to-end program repair with llms}. In \bibinfo{booktitle}{\emph{Proceedings of the 31st ACM Joint European Software Engineering Conference and Symposium on the Foundations of Software Engineering}}. \bibinfo{pages}{1646--1656}.
\newblock


\bibitem[Joshi et~al\mbox{.}(2023)]%
        {joshi2023repair}
\bibfield{author}{\bibinfo{person}{Harshit Joshi}, \bibinfo{person}{Jos{\'e}~Cambronero Sanchez}, \bibinfo{person}{Sumit Gulwani}, \bibinfo{person}{Vu Le}, \bibinfo{person}{Gust Verbruggen}, {and} \bibinfo{person}{Ivan Radi{\v{c}}ek}.} \bibinfo{year}{2023}\natexlab{}.
\newblock \showarticletitle{Repair is nearly generation: Multilingual program repair with llms}. In \bibinfo{booktitle}{\emph{Proceedings of the AAAI Conference on Artificial Intelligence}}, Vol.~\bibinfo{volume}{37}. \bibinfo{pages}{5131--5140}.
\newblock


\bibitem[Kim et~al\mbox{.}(2013)]%
        {kim2013automatic}
\bibfield{author}{\bibinfo{person}{Dongsun Kim}, \bibinfo{person}{Jaechang Nam}, \bibinfo{person}{Jaewoo Song}, {and} \bibinfo{person}{Sunghun Kim}.} \bibinfo{year}{2013}\natexlab{}.
\newblock \showarticletitle{Automatic patch generation learned from human-written patches}. In \bibinfo{booktitle}{\emph{2013 35th international conference on software engineering (ICSE)}}. IEEE, \bibinfo{pages}{802--811}.
\newblock


\bibitem[K{\"o}pf et~al\mbox{.}(2023)]%
        {kopf2023openassistant}
\bibfield{author}{\bibinfo{person}{Andreas K{\"o}pf}, \bibinfo{person}{Yannic Kilcher}, \bibinfo{person}{Dimitri Von~R{\"u}tte}, \bibinfo{person}{Sotiris Anagnostidis}, \bibinfo{person}{Zhi~Rui Tam}, \bibinfo{person}{Keith Stevens}, \bibinfo{person}{Abdullah Barhoum}, \bibinfo{person}{Duc Nguyen}, \bibinfo{person}{Oliver Stanley}, \bibinfo{person}{Rich{\'a}rd Nagyfi}, {et~al\mbox{.}}} \bibinfo{year}{2023}\natexlab{}.
\newblock \showarticletitle{Openassistant conversations-democratizing large language model alignment}.
\newblock \bibinfo{journal}{\emph{Advances in neural information processing systems}}  \bibinfo{volume}{36} (\bibinfo{year}{2023}), \bibinfo{pages}{47669--47681}.
\newblock


\bibitem[LaBash et~al\mbox{.}(2024)]%
        {labash2024res}
\bibfield{author}{\bibinfo{person}{Beck LaBash}, \bibinfo{person}{August Rosedale}, \bibinfo{person}{Alex Reents}, \bibinfo{person}{Lucas Negritto}, {and} \bibinfo{person}{Colin Wiel}.} \bibinfo{year}{2024}\natexlab{}.
\newblock \showarticletitle{Res-q: Evaluating code-editing large language model systems at the repository scale}.
\newblock \bibinfo{journal}{\emph{arXiv preprint arXiv:2406.16801}} (\bibinfo{year}{2024}).
\newblock


\bibitem[Li et~al\mbox{.}(2024)]%
        {li2024synthetic}
\bibfield{author}{\bibinfo{person}{Haoran Li}, \bibinfo{person}{Qingxiu Dong}, \bibinfo{person}{Zhengyang Tang}, \bibinfo{person}{Chaojun Wang}, \bibinfo{person}{Xingxing Zhang}, \bibinfo{person}{Haoyang Huang}, \bibinfo{person}{Shaohan Huang}, \bibinfo{person}{Xiaolong Huang}, \bibinfo{person}{Zeqiang Huang}, \bibinfo{person}{Dongdong Zhang}, {et~al\mbox{.}}} \bibinfo{year}{2024}\natexlab{}.
\newblock \showarticletitle{Synthetic data (almost) from scratch: Generalized instruction tuning for language models}.
\newblock \bibinfo{journal}{\emph{arXiv preprint arXiv:2402.13064}} (\bibinfo{year}{2024}).
\newblock


\bibitem[Li et~al\mbox{.}(2023b)]%
        {li2023codeeditor}
\bibfield{author}{\bibinfo{person}{Jia Li}, \bibinfo{person}{Ge Li}, \bibinfo{person}{Zhuo Li}, \bibinfo{person}{Zhi Jin}, \bibinfo{person}{Xing Hu}, \bibinfo{person}{Kechi Zhang}, {and} \bibinfo{person}{Zhiyi Fu}.} \bibinfo{year}{2023}\natexlab{b}.
\newblock \showarticletitle{Codeeditor: Learning to edit source code with pre-trained models}.
\newblock \bibinfo{journal}{\emph{ACM Transactions on Software Engineering and Methodology}} \bibinfo{volume}{32}, \bibinfo{number}{6} (\bibinfo{year}{2023}), \bibinfo{pages}{1--22}.
\newblock


\bibitem[Li et~al\mbox{.}(2023a)]%
        {li2023starcoder}
\bibfield{author}{\bibinfo{person}{Raymond Li}, \bibinfo{person}{Loubna~Ben Allal}, \bibinfo{person}{Yangtian Zi}, \bibinfo{person}{Niklas Muennighoff}, \bibinfo{person}{Denis Kocetkov}, \bibinfo{person}{Chenghao Mou}, \bibinfo{person}{Marc Marone}, \bibinfo{person}{Christopher Akiki}, \bibinfo{person}{Jia Li}, \bibinfo{person}{Jenny Chim}, {et~al\mbox{.}}} \bibinfo{year}{2023}\natexlab{a}.
\newblock \showarticletitle{Starcoder: may the source be with you!}
\newblock \bibinfo{journal}{\emph{arXiv preprint arXiv:2305.06161}} (\bibinfo{year}{2023}).
\newblock


\bibitem[Li et~al\mbox{.}(2025a)]%
        {li2025editlord}
\bibfield{author}{\bibinfo{person}{Weichen Li}, \bibinfo{person}{Albert Jan}, \bibinfo{person}{Baishakhi Ray}, \bibinfo{person}{Junfeng Yang}, \bibinfo{person}{Chengzhi Mao}, {and} \bibinfo{person}{Kexin Pei}.} \bibinfo{year}{2025}\natexlab{a}.
\newblock \showarticletitle{EDITLORD: Learning Code Transformation Rules for Code Editing}.
\newblock \bibinfo{journal}{\emph{arXiv preprint arXiv:2504.15284}} (\bibinfo{year}{2025}).
\newblock


\bibitem[Li et~al\mbox{.}(2025b)]%
        {li2025fea}
\bibfield{author}{\bibinfo{person}{Wei Li}, \bibinfo{person}{Xin Zhang}, \bibinfo{person}{Zhongxin Guo}, \bibinfo{person}{Shaoguang Mao}, \bibinfo{person}{Wen Luo}, \bibinfo{person}{Guangyue Peng}, \bibinfo{person}{Yangyu Huang}, \bibinfo{person}{Houfeng Wang}, {and} \bibinfo{person}{Scarlett Li}.} \bibinfo{year}{2025}\natexlab{b}.
\newblock \showarticletitle{Fea-bench: A benchmark for evaluating repository-level code generation for feature implementation}.
\newblock \bibinfo{journal}{\emph{arXiv preprint arXiv:2503.06680}} (\bibinfo{year}{2025}).
\newblock


\bibitem[Li et~al\mbox{.}(2022a)]%
        {li2022competition}
\bibfield{author}{\bibinfo{person}{Yujia Li}, \bibinfo{person}{David Choi}, \bibinfo{person}{Junyoung Chung}, \bibinfo{person}{Nate Kushman}, \bibinfo{person}{Julian Schrittwieser}, \bibinfo{person}{R{\'e}mi Leblond}, \bibinfo{person}{Tom Eccles}, \bibinfo{person}{James Keeling}, \bibinfo{person}{Felix Gimeno}, \bibinfo{person}{Agustin Dal~Lago}, {et~al\mbox{.}}} \bibinfo{year}{2022}\natexlab{a}.
\newblock \showarticletitle{Competition-level code generation with alphacode}.
\newblock \bibinfo{journal}{\emph{Science}} \bibinfo{volume}{378}, \bibinfo{number}{6624} (\bibinfo{year}{2022}), \bibinfo{pages}{1092--1097}.
\newblock


\bibitem[Li et~al\mbox{.}(2022b)]%
        {li2022automating}
\bibfield{author}{\bibinfo{person}{Zhiyu Li}, \bibinfo{person}{Shuai Lu}, \bibinfo{person}{Daya Guo}, \bibinfo{person}{Nan Duan}, \bibinfo{person}{Shailesh Jannu}, \bibinfo{person}{Grant Jenks}, \bibinfo{person}{Deep Majumder}, \bibinfo{person}{Jared Green}, \bibinfo{person}{Alexey Svyatkovskiy}, \bibinfo{person}{Shengyu Fu}, {et~al\mbox{.}}} \bibinfo{year}{2022}\natexlab{b}.
\newblock \showarticletitle{Automating code review activities by large-scale pre-training}. In \bibinfo{booktitle}{\emph{Proceedings of the 30th ACM Joint European Software Engineering Conference and Symposium on the Foundations of Software Engineering}}. \bibinfo{pages}{1035--1047}.
\newblock


\bibitem[Lin et~al\mbox{.}(2023)]%
        {lin2023cct5}
\bibfield{author}{\bibinfo{person}{Bo Lin}, \bibinfo{person}{Shangwen Wang}, \bibinfo{person}{Zhongxin Liu}, \bibinfo{person}{Yepang Liu}, \bibinfo{person}{Xin Xia}, {and} \bibinfo{person}{Xiaoguang Mao}.} \bibinfo{year}{2023}\natexlab{}.
\newblock \showarticletitle{Cct5: A code-change-oriented pre-trained model}. In \bibinfo{booktitle}{\emph{Proceedings of the 31st ACM Joint European Software Engineering Conference and Symposium on the Foundations of Software Engineering}}. \bibinfo{pages}{1509--1521}.
\newblock


\bibitem[Liu et~al\mbox{.}(2024)]%
        {liu2024coedpilot}
\bibfield{author}{\bibinfo{person}{Chenyan Liu}, \bibinfo{person}{Yufan Cai}, \bibinfo{person}{Yun Lin}, \bibinfo{person}{Yuhuan Huang}, \bibinfo{person}{Yunrui Pei}, \bibinfo{person}{Bo Jiang}, \bibinfo{person}{Ping Yang}, \bibinfo{person}{Jin~Song Dong}, {and} \bibinfo{person}{Hong Mei}.} \bibinfo{year}{2024}\natexlab{}.
\newblock \showarticletitle{CoEdPilot: Recommending Code Edits with Learned Prior Edit Relevance, Project-wise Awareness, and Interactive Nature}. In \bibinfo{booktitle}{\emph{Proceedings of the 33rd ACM SIGSOFT International Symposium on Software Testing and Analysis}}. \bibinfo{pages}{466--478}.
\newblock


\bibitem[Lozhkov et~al\mbox{.}(2024)]%
        {lozhkov2024starcoder}
\bibfield{author}{\bibinfo{person}{Anton Lozhkov}, \bibinfo{person}{Raymond Li}, \bibinfo{person}{Loubna~Ben Allal}, \bibinfo{person}{Federico Cassano}, \bibinfo{person}{Joel Lamy-Poirier}, \bibinfo{person}{Nouamane Tazi}, \bibinfo{person}{Ao Tang}, \bibinfo{person}{Dmytro Pykhtar}, \bibinfo{person}{Jiawei Liu}, \bibinfo{person}{Yuxiang Wei}, {et~al\mbox{.}}} \bibinfo{year}{2024}\natexlab{}.
\newblock \showarticletitle{Starcoder 2 and the stack v2: The next generation}.
\newblock \bibinfo{journal}{\emph{arXiv preprint arXiv:2402.19173}} (\bibinfo{year}{2024}).
\newblock


\bibitem[Luo et~al\mbox{.}(2024)]%
        {luo2024wizardcoder}
\bibfield{author}{\bibinfo{person}{Ziyang Luo}, \bibinfo{person}{Can Xu}, \bibinfo{person}{Pu Zhao}, \bibinfo{person}{Qingfeng Sun}, \bibinfo{person}{Xiubo Geng}, \bibinfo{person}{Wenxiang Hu}, \bibinfo{person}{Chongyang Tao}, \bibinfo{person}{Jing Ma}, \bibinfo{person}{Qingwei Lin}, {and} \bibinfo{person}{Daxin Jiang}.} \bibinfo{year}{2024}\natexlab{}.
\newblock \showarticletitle{WizardCoder: Empowering Code Large Language Models with Evol-Instruct}. In \bibinfo{booktitle}{\emph{ICLR}}.
\newblock


\bibitem[Mina et~al\mbox{.}(2025)]%
        {mina2025cognitive}
\bibfield{author}{\bibinfo{person}{Mario Mina}, \bibinfo{person}{Valle Ruiz-Fern{\'a}ndez}, \bibinfo{person}{J{\'u}lia Falc{\~a}o}, \bibinfo{person}{Luis Vasquez-Reina}, {and} \bibinfo{person}{Aitor Gonzalez-Agirre}.} \bibinfo{year}{2025}\natexlab{}.
\newblock \showarticletitle{Cognitive Biases, Task Complexity, and Result Interpretability in Large Language Models}. In \bibinfo{booktitle}{\emph{Proceedings of the 31st International Conference on Computational Linguistics}}, \bibfield{editor}{\bibinfo{person}{Owen Rambow}, \bibinfo{person}{Leo Wanner}, \bibinfo{person}{Marianna Apidianaki}, \bibinfo{person}{Hend Al-Khalifa}, \bibinfo{person}{Barbara~Di Eugenio}, {and} \bibinfo{person}{Steven Schockaert}} (Eds.). \bibinfo{publisher}{Association for Computational Linguistics}, \bibinfo{address}{Abu Dhabi, UAE}, \bibinfo{pages}{1767--1784}.
\newblock
\urldef\tempurl%
\url{https://aclanthology.org/2025.coling-main.120/}
\showURL{%
\tempurl}


\bibitem[Moon et~al\mbox{.}(2023)]%
        {moon2023coffee}
\bibfield{author}{\bibinfo{person}{Seungjun Moon}, \bibinfo{person}{Hyungjoo Chae}, \bibinfo{person}{Yongho Song}, \bibinfo{person}{Taeyoon Kwon}, \bibinfo{person}{Dongjin Kang}, \bibinfo{person}{Kai Tzu-iunn Ong}, \bibinfo{person}{Seung-won Hwang}, {and} \bibinfo{person}{Jinyoung Yeo}.} \bibinfo{year}{2023}\natexlab{}.
\newblock \showarticletitle{Coffee: Boost your code llms by fixing bugs with feedback}.
\newblock \bibinfo{journal}{\emph{arXiv preprint arXiv:2311.07215}} (\bibinfo{year}{2023}).
\newblock


\bibitem[Mozannar et~al\mbox{.}(2024)]%
        {mozannar2024reading}
\bibfield{author}{\bibinfo{person}{Hussein Mozannar}, \bibinfo{person}{Gagan Bansal}, \bibinfo{person}{Adam Fourney}, {and} \bibinfo{person}{Eric Horvitz}.} \bibinfo{year}{2024}\natexlab{}.
\newblock \showarticletitle{Reading between the lines: Modeling user behavior and costs in AI-assisted programming}. In \bibinfo{booktitle}{\emph{Proceedings of the 2024 CHI Conference on Human Factors in Computing Systems}}. \bibinfo{pages}{1--16}.
\newblock


\bibitem[Muennighoff et~al\mbox{.}(2023)]%
        {muennighoff2023octopack}
\bibfield{author}{\bibinfo{person}{Niklas Muennighoff}, \bibinfo{person}{Qian Liu}, \bibinfo{person}{Armel Zebaze}, \bibinfo{person}{Qinkai Zheng}, \bibinfo{person}{Binyuan Hui}, \bibinfo{person}{Terry~Yue Zhuo}, \bibinfo{person}{Swayam Singh}, \bibinfo{person}{Xiangru Tang}, \bibinfo{person}{Leandro Von~Werra}, {and} \bibinfo{person}{Shayne Longpre}.} \bibinfo{year}{2023}\natexlab{}.
\newblock \showarticletitle{Octopack: Instruction tuning code large language models}. In \bibinfo{booktitle}{\emph{NeurIPS 2023 Workshop on Instruction Tuning and Instruction Following}}.
\newblock


\bibitem[Nguyen et~al\mbox{.}(2016)]%
        {nguyen2016api}
\bibfield{author}{\bibinfo{person}{Anh~Tuan Nguyen}, \bibinfo{person}{Michael Hilton}, \bibinfo{person}{Mihai Codoban}, \bibinfo{person}{Hoan~Anh Nguyen}, \bibinfo{person}{Lily Mast}, \bibinfo{person}{Eli Rademacher}, \bibinfo{person}{Tien~N Nguyen}, {and} \bibinfo{person}{Danny Dig}.} \bibinfo{year}{2016}\natexlab{}.
\newblock \showarticletitle{API code recommendation using statistical learning from fine-grained changes}. In \bibinfo{booktitle}{\emph{Proceedings of the 2016 24th ACM SIGSOFT International Symposium on Foundations of Software Engineering}}. \bibinfo{pages}{511--522}.
\newblock


\bibitem[Nguyen et~al\mbox{.}(2013)]%
        {nguyen2013study}
\bibfield{author}{\bibinfo{person}{Hoan~Anh Nguyen}, \bibinfo{person}{Anh~Tuan Nguyen}, \bibinfo{person}{Tung~Thanh Nguyen}, \bibinfo{person}{Tien~N Nguyen}, {and} \bibinfo{person}{Hridesh Rajan}.} \bibinfo{year}{2013}\natexlab{}.
\newblock \showarticletitle{A study of repetitiveness of code changes in software evolution}. In \bibinfo{booktitle}{\emph{2013 28th IEEE/ACM International Conference on Automated Software Engineering (ASE)}}. IEEE, \bibinfo{pages}{180--190}.
\newblock


\bibitem[Nguyen et~al\mbox{.}(2010)]%
        {nguyen2010graph}
\bibfield{author}{\bibinfo{person}{Hoan~Anh Nguyen}, \bibinfo{person}{Tung~Thanh Nguyen}, \bibinfo{person}{Gary Wilson~Jr}, \bibinfo{person}{Anh~Tuan Nguyen}, \bibinfo{person}{Miryung Kim}, {and} \bibinfo{person}{Tien~N Nguyen}.} \bibinfo{year}{2010}\natexlab{}.
\newblock \showarticletitle{A graph-based approach to API usage adaptation}.
\newblock \bibinfo{journal}{\emph{ACM Sigplan Notices}} \bibinfo{volume}{45}, \bibinfo{number}{10} (\bibinfo{year}{2010}), \bibinfo{pages}{302--321}.
\newblock


\bibitem[Olausson et~al\mbox{.}(2023)]%
        {olausson2023self}
\bibfield{author}{\bibinfo{person}{Theo~X Olausson}, \bibinfo{person}{Jeevana~Priya Inala}, \bibinfo{person}{Chenglong Wang}, \bibinfo{person}{Jianfeng Gao}, {and} \bibinfo{person}{Armando Solar-Lezama}.} \bibinfo{year}{2023}\natexlab{}.
\newblock \showarticletitle{Is self-repair a silver bullet for code generation?}
\newblock \bibinfo{journal}{\emph{arXiv preprint arXiv:2306.09896}} (\bibinfo{year}{2023}).
\newblock


\bibitem[OpenAI(2023a)]%
        {openai2023gpt4}
\bibfield{author}{\bibinfo{person}{OpenAI}.} \bibinfo{year}{2023}\natexlab{a}.
\newblock \bibinfo{booktitle}{\emph{GPT-4 Technical Report}}.
\newblock \bibinfo{type}{Technical report}. \bibinfo{institution}{OpenAI}.
\newblock
\urldef\tempurl%
\url{https://cdn.openai.com/papers/gpt-4.pdf}
\showURL{%
\tempurl}


\bibitem[OpenAI(2023b)]%
        {openai2023chatgptapi}
\bibfield{author}{\bibinfo{person}{OpenAI}.} \bibinfo{year}{2023}\natexlab{b}.
\newblock \bibinfo{title}{Introducing ChatGPT and Whisper APIs}.
\newblock \bibinfo{howpublished}{\url{https://openai.com/blog/introducing-chatgpt-and-whisper-apis}}.
\newblock


\bibitem[OpenAI(2023c)]%
        {openai-api}
\bibfield{author}{\bibinfo{person}{OpenAI}.} \bibinfo{year}{2023}\natexlab{c}.
\newblock \bibinfo{booktitle}{\emph{OpenAI API}}.
\newblock
\urldef\tempurl%
\url{https://platform.openai.com}
\showURL{%
\tempurl}


\bibitem[Ouyang et~al\mbox{.}(2025)]%
        {ouyang2025kernelbench}
\bibfield{author}{\bibinfo{person}{Anne Ouyang}, \bibinfo{person}{Simon Guo}, \bibinfo{person}{Simran Arora}, \bibinfo{person}{Alex~L Zhang}, \bibinfo{person}{William Hu}, \bibinfo{person}{Christopher R{\'e}}, {and} \bibinfo{person}{Azalia Mirhoseini}.} \bibinfo{year}{2025}\natexlab{}.
\newblock \showarticletitle{Kernelbench: Can llms write efficient gpu kernels?}
\newblock \bibinfo{journal}{\emph{arXiv preprint arXiv:2502.10517}} (\bibinfo{year}{2025}).
\newblock


\bibitem[{Python Software Foundation}(2024)]%
        {python-difflib}
\bibfield{author}{\bibinfo{person}{{Python Software Foundation}}.} \bibinfo{year}{2024}\natexlab{}.
\newblock \bibinfo{booktitle}{\emph{difflib --- Helpers for computing deltas}}.
\newblock
\urldef\tempurl%
\url{https://docs.python.org/3.13/library/difflib.html}
\showURL{%
\tempurl}
\newblock
\shownote{Python 3.13 documentation}.


\bibitem[Roziere et~al\mbox{.}(2023)]%
        {roziere2023code}
\bibfield{author}{\bibinfo{person}{Baptiste Roziere}, \bibinfo{person}{Jonas Gehring}, \bibinfo{person}{Fabian Gloeckle}, \bibinfo{person}{Sten Sootla}, \bibinfo{person}{Itai Gat}, \bibinfo{person}{Xiaoqing~Ellen Tan}, \bibinfo{person}{Yossi Adi}, \bibinfo{person}{Jingyu Liu}, \bibinfo{person}{Romain Sauvestre}, \bibinfo{person}{Tal Remez}, {et~al\mbox{.}}} \bibinfo{year}{2023}\natexlab{}.
\newblock \showarticletitle{Code llama: Open foundation models for code}.
\newblock \bibinfo{journal}{\emph{arXiv preprint arXiv:2308.12950}} (\bibinfo{year}{2023}).
\newblock


\bibitem[Shen et~al\mbox{.}(2024)]%
        {shen2024efficient}
\bibfield{author}{\bibinfo{person}{Li Shen}, \bibinfo{person}{Yan Sun}, \bibinfo{person}{Zhiyuan Yu}, \bibinfo{person}{Liang Ding}, \bibinfo{person}{Xinmei Tian}, {and} \bibinfo{person}{Dacheng Tao}.} \bibinfo{year}{2024}\natexlab{}.
\newblock \showarticletitle{On efficient training of large-scale deep learning models}.
\newblock \bibinfo{journal}{\emph{Comput. Surveys}} \bibinfo{volume}{57}, \bibinfo{number}{3} (\bibinfo{year}{2024}), \bibinfo{pages}{1--36}.
\newblock


\bibitem[Shinn et~al\mbox{.}(2023)]%
        {shinn2023reflexion}
\bibfield{author}{\bibinfo{person}{Noah Shinn}, \bibinfo{person}{Federico Cassano}, \bibinfo{person}{Ashwin Gopinath}, \bibinfo{person}{Karthik Narasimhan}, {and} \bibinfo{person}{Shunyu Yao}.} \bibinfo{year}{2023}\natexlab{}.
\newblock \showarticletitle{Reflexion: Language agents with verbal reinforcement learning}.
\newblock \bibinfo{journal}{\emph{Advances in Neural Information Processing Systems}}  \bibinfo{volume}{36} (\bibinfo{year}{2023}), \bibinfo{pages}{8634--8652}.
\newblock


\bibitem[Shirafuji et~al\mbox{.}(2023)]%
        {shirafuji2023refactoring}
\bibfield{author}{\bibinfo{person}{Atsushi Shirafuji}, \bibinfo{person}{Yusuke Oda}, \bibinfo{person}{Jun Suzuki}, \bibinfo{person}{Makoto Morishita}, {and} \bibinfo{person}{Yutaka Watanobe}.} \bibinfo{year}{2023}\natexlab{}.
\newblock \showarticletitle{Refactoring programs using large language models with few-shot examples}. In \bibinfo{booktitle}{\emph{2023 30th Asia-Pacific Software Engineering Conference (APSEC)}}. IEEE, \bibinfo{pages}{151--160}.
\newblock


\bibitem[Shypula et~al\mbox{.}(2023)]%
        {shypula2023learning}
\bibfield{author}{\bibinfo{person}{Alexander Shypula}, \bibinfo{person}{Aman Madaan}, \bibinfo{person}{Yimeng Zeng}, \bibinfo{person}{Uri Alon}, \bibinfo{person}{Jacob Gardner}, \bibinfo{person}{Milad Hashemi}, \bibinfo{person}{Graham Neubig}, \bibinfo{person}{Parthasarathy Ranganathan}, \bibinfo{person}{Osbert Bastani}, {and} \bibinfo{person}{Amir Yazdanbakhsh}.} \bibinfo{year}{2023}\natexlab{}.
\newblock \showarticletitle{Learning performance-improving code edits}.
\newblock \bibinfo{journal}{\emph{arXiv preprint arXiv:2302.07867}} (\bibinfo{year}{2023}).
\newblock


\bibitem[Silva et~al\mbox{.}(2021)]%
        {silva2021topic}
\bibfield{author}{\bibinfo{person}{Camila~Costa Silva}, \bibinfo{person}{Matthias Galster}, {and} \bibinfo{person}{Fabian Gilson}.} \bibinfo{year}{2021}\natexlab{}.
\newblock \showarticletitle{Topic modeling in software engineering research}.
\newblock \bibinfo{journal}{\emph{Empirical Software Engineering}} \bibinfo{volume}{26}, \bibinfo{number}{6} (\bibinfo{year}{2021}), \bibinfo{pages}{120}.
\newblock


\bibitem[Singhal et~al\mbox{.}(2024)]%
        {singhal2024nofuneval}
\bibfield{author}{\bibinfo{person}{Manav Singhal}, \bibinfo{person}{Tushar Aggarwal}, \bibinfo{person}{Abhijeet Awasthi}, \bibinfo{person}{Nagarajan Natarajan}, {and} \bibinfo{person}{Aditya Kanade}.} \bibinfo{year}{2024}\natexlab{}.
\newblock \showarticletitle{Nofuneval: Funny how code lms falter on requirements beyond functional correctness}.
\newblock \bibinfo{journal}{\emph{arXiv preprint arXiv:2401.15963}} (\bibinfo{year}{2024}).
\newblock


\bibitem[Taori et~al\mbox{.}(2023)]%
        {taori2023stanford}
\bibfield{author}{\bibinfo{person}{Rohan Taori}, \bibinfo{person}{Ishaan Gulrajani}, \bibinfo{person}{Tianyi Zhang}, \bibinfo{person}{Yann Dubois}, \bibinfo{person}{Xuechen Li}, \bibinfo{person}{Carlos Guestrin}, \bibinfo{person}{Percy Liang}, {and} \bibinfo{person}{Tatsunori~B Hashimoto}.} \bibinfo{year}{2023}\natexlab{}.
\newblock \bibinfo{title}{Stanford alpaca: An instruction-following llama model}.
\newblock


\bibitem[Tufano et~al\mbox{.}(2019)]%
        {tufano2019learning}
\bibfield{author}{\bibinfo{person}{Michele Tufano}, \bibinfo{person}{Jevgenija Pantiuchina}, \bibinfo{person}{Cody Watson}, \bibinfo{person}{Gabriele Bavota}, {and} \bibinfo{person}{Denys Poshyvanyk}.} \bibinfo{year}{2019}\natexlab{}.
\newblock \showarticletitle{On learning meaningful code changes via neural machine translation}. In \bibinfo{booktitle}{\emph{2019 IEEE/ACM 41st International Conference on Software Engineering (ICSE)}}. IEEE, \bibinfo{pages}{25--36}.
\newblock


\bibitem[Wang et~al\mbox{.}(2023)]%
        {wang-etal-2023-self-instruct}
\bibfield{author}{\bibinfo{person}{Yizhong Wang}, \bibinfo{person}{Yeganeh Kordi}, \bibinfo{person}{Swaroop Mishra}, \bibinfo{person}{Alisa Liu}, \bibinfo{person}{Noah~A. Smith}, \bibinfo{person}{Daniel Khashabi}, {and} \bibinfo{person}{Hannaneh Hajishirzi}.} \bibinfo{year}{2023}\natexlab{}.
\newblock \showarticletitle{Self-Instruct: Aligning Language Models with Self-Generated Instructions}. In \bibinfo{booktitle}{\emph{Proceedings of the 61st Annual Meeting of the Association for Computational Linguistics (Volume 1: Long Papers)}}, \bibfield{editor}{\bibinfo{person}{Anna Rogers}, \bibinfo{person}{Jordan Boyd-Graber}, {and} \bibinfo{person}{Naoaki Okazaki}} (Eds.). \bibinfo{publisher}{Association for Computational Linguistics}, \bibinfo{address}{Toronto, Canada}, \bibinfo{pages}{13484--13508}.
\newblock
\href{https://doi.org/10.18653/v1/2023.acl-long.754}{doi:\nolinkurl{10.18653/v1/2023.acl-long.754}}


\bibitem[Wang et~al\mbox{.}(2024)]%
        {wang2024codeclm}
\bibfield{author}{\bibinfo{person}{Zifeng Wang}, \bibinfo{person}{Chun-Liang Li}, \bibinfo{person}{Vincent Perot}, \bibinfo{person}{Long Le}, \bibinfo{person}{Jin Miao}, \bibinfo{person}{Zizhao Zhang}, \bibinfo{person}{Chen-Yu Lee}, {and} \bibinfo{person}{Tomas Pfister}.} \bibinfo{year}{2024}\natexlab{}.
\newblock \showarticletitle{{C}odec{LM}: Aligning Language Models with Tailored Synthetic Data}. In \bibinfo{booktitle}{\emph{Findings of the Association for Computational Linguistics: NAACL 2024}}, \bibfield{editor}{\bibinfo{person}{Kevin Duh}, \bibinfo{person}{Helena Gomez}, {and} \bibinfo{person}{Steven Bethard}} (Eds.). \bibinfo{publisher}{Association for Computational Linguistics}, \bibinfo{address}{Mexico City, Mexico}, \bibinfo{pages}{3712--3729}.
\newblock
\href{https://doi.org/10.18653/v1/2024.findings-naacl.235}{doi:\nolinkurl{10.18653/v1/2024.findings-naacl.235}}


\bibitem[Wei et~al\mbox{.}(2024c)]%
        {wei2024improving}
\bibfield{author}{\bibinfo{person}{Anjiang Wei}, \bibinfo{person}{Allen Nie}, \bibinfo{person}{Thiago~SFX Teixeira}, \bibinfo{person}{Rohan Yadav}, \bibinfo{person}{Wonchan Lee}, \bibinfo{person}{Ke Wang}, {and} \bibinfo{person}{Alex Aiken}.} \bibinfo{year}{2024}\natexlab{c}.
\newblock \showarticletitle{Improving Parallel Program Performance with LLM Optimizers via Agent-System Interfaces}.
\newblock \bibinfo{journal}{\emph{arXiv preprint arXiv:2410.15625}} (\bibinfo{year}{2024}).
\newblock


\bibitem[Wei et~al\mbox{.}(2024b)]%
        {wei2024coeditor}
\bibfield{author}{\bibinfo{person}{Jiayi Wei}, \bibinfo{person}{Greg Durrett}, {and} \bibinfo{person}{Isil Dillig}.} \bibinfo{year}{2024}\natexlab{b}.
\newblock \showarticletitle{Coeditor: Leveraging Repo-level Diffs for Code Auto-editing}. In \bibinfo{booktitle}{\emph{The Twelfth International Conference on Learning Representations}}.
\newblock


\bibitem[Wei et~al\mbox{.}(2024a)]%
        {wei2024selfcodealign}
\bibfield{author}{\bibinfo{person}{Yuxiang Wei}, \bibinfo{person}{Federico Cassano}, \bibinfo{person}{Jiawei Liu}, \bibinfo{person}{Yifeng Ding}, \bibinfo{person}{Naman Jain}, \bibinfo{person}{Zachary Mueller}, \bibinfo{person}{Harm de Vries}, \bibinfo{person}{Leandro Von~Werra}, \bibinfo{person}{Arjun Guha}, {and} \bibinfo{person}{LINGMING ZHANG}.} \bibinfo{year}{2024}\natexlab{a}.
\newblock \showarticletitle{SelfCodeAlign: Self-Alignment for Code Generation}. In \bibinfo{booktitle}{\emph{The Thirty-eighth Annual Conference on Neural Information Processing Systems}}.
\newblock


\bibitem[Wei et~al\mbox{.}(2024d)]%
        {wei2024magicoder}
\bibfield{author}{\bibinfo{person}{Yuxiang Wei}, \bibinfo{person}{Zhe Wang}, \bibinfo{person}{Jiawei Liu}, \bibinfo{person}{Yifeng Ding}, {and} \bibinfo{person}{Lingming Zhang}.} \bibinfo{year}{2024}\natexlab{d}.
\newblock \showarticletitle{Magicoder: empowering code generation with OSS-INSTRUCT}. In \bibinfo{booktitle}{\emph{Proceedings of the 41st International Conference on Machine Learning}} (Vienna, Austria) \emph{(\bibinfo{series}{ICML'24})}. \bibinfo{publisher}{JMLR.org}, Article \bibinfo{articleno}{2158}, \bibinfo{numpages}{26}~pages.
\newblock


\bibitem[Wei et~al\mbox{.}(2023)]%
        {wei2023copiloting}
\bibfield{author}{\bibinfo{person}{Yuxiang Wei}, \bibinfo{person}{Chunqiu~Steven Xia}, {and} \bibinfo{person}{Lingming Zhang}.} \bibinfo{year}{2023}\natexlab{}.
\newblock \showarticletitle{Copiloting the copilots: Fusing large language models with completion engines for automated program repair}. In \bibinfo{booktitle}{\emph{Proceedings of the 31st ACM Joint European Software Engineering Conference and Symposium on the Foundations of Software Engineering}}. \bibinfo{pages}{172--184}.
\newblock


\bibitem[Xia and Zhang(2024)]%
        {xia2024automated}
\bibfield{author}{\bibinfo{person}{Chunqiu~Steven Xia} {and} \bibinfo{person}{Lingming Zhang}.} \bibinfo{year}{2024}\natexlab{}.
\newblock \showarticletitle{Automated program repair via conversation: Fixing 162 out of 337 bugs for \$0.42 each using ChatGPT}. In \bibinfo{booktitle}{\emph{Proceedings of the 33rd ACM SIGSOFT International Symposium on Software Testing and Analysis}}. \bibinfo{pages}{819--831}.
\newblock


\bibitem[Xie et~al\mbox{.}(2025)]%
        {xie2025swe}
\bibfield{author}{\bibinfo{person}{Chengxing Xie}, \bibinfo{person}{Bowen Li}, \bibinfo{person}{Chang Gao}, \bibinfo{person}{He Du}, \bibinfo{person}{Wai Lam}, \bibinfo{person}{Difan Zou}, {and} \bibinfo{person}{Kai Chen}.} \bibinfo{year}{2025}\natexlab{}.
\newblock \showarticletitle{Swe-fixer: Training open-source llms for effective and efficient github issue resolution}.
\newblock \bibinfo{journal}{\emph{arXiv preprint arXiv:2501.05040}} (\bibinfo{year}{2025}).
\newblock


\bibitem[Xu et~al\mbox{.}(2024)]%
        {xu2024wizardlm}
\bibfield{author}{\bibinfo{person}{Can Xu}, \bibinfo{person}{Qingfeng Sun}, \bibinfo{person}{Kai Zheng}, \bibinfo{person}{Xiubo Geng}, \bibinfo{person}{Pu Zhao}, \bibinfo{person}{Jiazhan Feng}, \bibinfo{person}{Chongyang Tao}, \bibinfo{person}{Qingwei Lin}, {and} \bibinfo{person}{Daxin Jiang}.} \bibinfo{year}{2024}\natexlab{}.
\newblock \showarticletitle{WizardLM: Empowering large pre-trained language models to follow complex instructions}. In \bibinfo{booktitle}{\emph{The Twelfth International Conference on Learning Representations}}.
\newblock


\bibitem[Yang et~al\mbox{.}(2025)]%
        {yang2025qwen3}
\bibfield{author}{\bibinfo{person}{An Yang}, \bibinfo{person}{Anfeng Li}, \bibinfo{person}{Baosong Yang}, \bibinfo{person}{Beichen Zhang}, \bibinfo{person}{Binyuan Hui}, \bibinfo{person}{Bo Zheng}, \bibinfo{person}{Bowen Yu}, \bibinfo{person}{Chang Gao}, \bibinfo{person}{Chengen Huang}, \bibinfo{person}{Chenxu Lv}, {et~al\mbox{.}}} \bibinfo{year}{2025}\natexlab{}.
\newblock \showarticletitle{Qwen3 technical report}.
\newblock \bibinfo{journal}{\emph{arXiv preprint arXiv:2505.09388}} (\bibinfo{year}{2025}).
\newblock


\bibitem[Yu et~al\mbox{.}(2023)]%
        {yu2023wavecoder}
\bibfield{author}{\bibinfo{person}{Zhaojian Yu}, \bibinfo{person}{Xin Zhang}, \bibinfo{person}{Ning Shang}, \bibinfo{person}{Yangyu Huang}, \bibinfo{person}{Can Xu}, \bibinfo{person}{Yishujie Zhao}, \bibinfo{person}{Wenxiang Hu}, {and} \bibinfo{person}{Qiufeng Yin}.} \bibinfo{year}{2023}\natexlab{}.
\newblock \showarticletitle{Wavecoder: Widespread and versatile enhancement for code large language models by instruction tuning}.
\newblock \bibinfo{journal}{\emph{arXiv preprint arXiv:2312.14187}} (\bibinfo{year}{2023}).
\newblock


\bibitem[Zhang et~al\mbox{.}(2023b)]%
        {zhang2023self}
\bibfield{author}{\bibinfo{person}{Kechi Zhang}, \bibinfo{person}{Zhuo Li}, \bibinfo{person}{Jia Li}, \bibinfo{person}{Ge Li}, {and} \bibinfo{person}{Zhi Jin}.} \bibinfo{year}{2023}\natexlab{b}.
\newblock \showarticletitle{Self-edit: Fault-aware code editor for code generation}.
\newblock \bibinfo{journal}{\emph{arXiv preprint arXiv:2305.04087}} (\bibinfo{year}{2023}).
\newblock


\bibitem[Zhang et~al\mbox{.}(2023a)]%
        {zhang2023survey}
\bibfield{author}{\bibinfo{person}{Quanjun Zhang}, \bibinfo{person}{Chunrong Fang}, \bibinfo{person}{Yang Xie}, \bibinfo{person}{Yaxin Zhang}, \bibinfo{person}{Yun Yang}, \bibinfo{person}{Weisong Sun}, \bibinfo{person}{Shengcheng Yu}, {and} \bibinfo{person}{Zhenyu Chen}.} \bibinfo{year}{2023}\natexlab{a}.
\newblock \showarticletitle{A survey on large language models for software engineering}.
\newblock \bibinfo{journal}{\emph{arXiv preprint arXiv:2312.15223}} (\bibinfo{year}{2023}).
\newblock


\bibitem[Zhao et~al\mbox{.}(2024)]%
        {zhao2024self}
\bibfield{author}{\bibinfo{person}{Chenyang Zhao}, \bibinfo{person}{Xueying Jia}, \bibinfo{person}{Vijay Viswanathan}, \bibinfo{person}{Tongshuang Wu}, {and} \bibinfo{person}{Graham Neubig}.} \bibinfo{year}{2024}\natexlab{}.
\newblock \showarticletitle{Self-guide: Better task-specific instruction following via self-synthetic finetuning}.
\newblock \bibinfo{journal}{\emph{arXiv preprint arXiv:2407.12874}} (\bibinfo{year}{2024}).
\newblock


\bibitem[Zheng et~al\mbox{.}(2024)]%
        {zheng2024llamafactory}
\bibfield{author}{\bibinfo{person}{Yaowei Zheng}, \bibinfo{person}{Richong Zhang}, \bibinfo{person}{Junhao Zhang}, \bibinfo{person}{Yanhan Ye}, \bibinfo{person}{Zheyan Luo}, \bibinfo{person}{Zhangchi Feng}, {and} \bibinfo{person}{Yongqiang Ma}.} \bibinfo{year}{2024}\natexlab{}.
\newblock \showarticletitle{LlamaFactory: Unified Efficient Fine-Tuning of 100+ Language Models}. In \bibinfo{booktitle}{\emph{Proceedings of the 62nd Annual Meeting of the Association for Computational Linguistics (Volume 3: System Demonstrations)}}. \bibinfo{publisher}{Association for Computational Linguistics}, \bibinfo{address}{Bangkok, Thailand}.
\newblock
\urldef\tempurl%
\url{http://arxiv.org/abs/2403.13372}
\showURL{%
\tempurl}


\bibitem[Zhou et~al\mbox{.}(2023b)]%
        {zhou2023hybrid}
\bibfield{author}{\bibinfo{person}{Bingzhe Zhou}, \bibinfo{person}{Xinying Wang}, \bibinfo{person}{Shengbin Xu}, \bibinfo{person}{Yuan Yao}, \bibinfo{person}{Minxue Pan}, \bibinfo{person}{Feng Xu}, {and} \bibinfo{person}{Xiaoxing Ma}.} \bibinfo{year}{2023}\natexlab{b}.
\newblock \showarticletitle{Hybrid api migration: A marriage of small api mapping models and large language models}. In \bibinfo{booktitle}{\emph{Proceedings of the 14th Asia-Pacific Symposium on Internetware}}. \bibinfo{pages}{12--21}.
\newblock


\bibitem[Zhou et~al\mbox{.}(2023a)]%
        {zhou2023lima}
\bibfield{author}{\bibinfo{person}{Chunting Zhou}, \bibinfo{person}{Pengfei Liu}, \bibinfo{person}{Puxin Xu}, \bibinfo{person}{Srinivasan Iyer}, \bibinfo{person}{Jiao Sun}, \bibinfo{person}{Yuning Mao}, \bibinfo{person}{Xuezhe Ma}, \bibinfo{person}{Avia Efrat}, \bibinfo{person}{Ping Yu}, \bibinfo{person}{Lili Yu}, {et~al\mbox{.}}} \bibinfo{year}{2023}\natexlab{a}.
\newblock \showarticletitle{Lima: Less is more for alignment}.
\newblock \bibinfo{journal}{\emph{Advances in Neural Information Processing Systems}}  \bibinfo{volume}{36} (\bibinfo{year}{2023}), \bibinfo{pages}{55006--55021}.
\newblock


\bibitem[Ziftci et~al\mbox{.}(2025)]%
        {ziftci2025migrating}
\bibfield{author}{\bibinfo{person}{Celal Ziftci}, \bibinfo{person}{Stoyan Nikolov}, \bibinfo{person}{Anna Sj{\"o}vall}, \bibinfo{person}{Bo Kim}, \bibinfo{person}{Daniele Codecasa}, {and} \bibinfo{person}{Max Kim}.} \bibinfo{year}{2025}\natexlab{}.
\newblock \showarticletitle{Migrating code at scale with llms at google}. In \bibinfo{booktitle}{\emph{Proceedings of the 33rd ACM International Conference on the Foundations of Software Engineering}}. \bibinfo{pages}{162--173}.
\newblock


\end{thebibliography}


\end{document}